\documentclass[superscriptaddress,11pt,nofootinbib]{revtex4}
\usepackage{amssymb,graphicx,amsmath,array,verbatim,setspace}

\newcommand{\beq}{\begin{eqnarray}}
\newcommand{\eeq}{\end{eqnarray}}
\newcommand{\p}{\partial}

\newcommand{\vs}[1]{\vspace{#1 mm}}
\newcommand{\hs}[1]{\hspace{#1 mm}}
\newcommand{\bpm}{\begin{pmatrix}}
\newcommand{\epm}{\end{pmatrix}}
\newcommand{\Z}{\mathbb{Z}}
\newcommand{\R}{\mathbb{R}}
\newcommand{\C}{\mathbb{C}}
\newcommand{\tr}{{\rm Tr}}
\newcommand{\D}{\mathcal D}

\newcommand{\ba}{\left(\begin{array}}
\newcommand{\ea}{\end{array} \right)}

\makeatletter
\@addtoreset{equation}{section}

\makeatother

\usepackage[usenames]{color}

\begin{document}
\title{Bion non-perturbative contributions versus infrared renormalons \\
in two-dimensional $\C P^{N-1}$ models
\vspace{5mm}}

\author{Toshiaki Fujimori}
\email{toshiaki.fujimori018(at)gmail.com}
\address{Department of Physics, and Research and 
Education Center for Natural Sciences, 
Keio University, 4-1-1 Hiyoshi, Yokohama, Kanagawa 223-8521, Japan}

\author{Syo Kamata}
\email{skamata(at)ncsu.edu}
\address{Department of Physics, North Carolina State University, Raleigh, NC 27695, USA}

\author{Tatsuhiro Misumi}
\email{misumi(at)phys.akita-u.ac.jp}
\address{Department of Mathematical Science, Akita 
University,  Akita 010-8502, Japan
}
\address{Department of Physics, and Research and 
Education Center for Natural Sciences, 
Keio University, 4-1-1 Hiyoshi, Yokohama, Kanagawa 223-8521, Japan}
\address{iTHEMS, RIKEN,
2-1 Hirasawa, Wako, Saitama 351-0198, Japan
}

\author{\\Muneto Nitta}
\email{nitta(at)phys-h.keio.ac.jp}
\address{Department of Physics, and Research and 
Education Center for Natural Sciences, 
Keio University, 4-1-1 Hiyoshi, Yokohama, Kanagawa 223-8521, Japan}

\author{Norisuke Sakai}
\email{norisuke.sakai(at)gmail.com}
\address{Department of Physics, and Research and 
Education Center for Natural Sciences, 
Keio University, 4-1-1 Hiyoshi, Yokohama, Kanagawa 223-8521, Japan}

\begin{abstract} \vs{10}
We derive the semiclassical contributions 
from the real and complex bions 
in the two-dimensional $\C P^{N-1}$ sigma model
on ${\mathbb R}\times S^{1}$ with a twisted boundary condition. 
The bion configurations are saddle points of 
the complexified Euclidean action, 
which can be viewed as bound states of 
a pair of fractional instantons 
with opposite topological charges. 
We first derive the bion solutions 
by solving the equation of motion
in the model with a potential which simulates 
an interaction induced by fermions 
in the $\C P^{N-1}$ quantum mechanics. 
The bion solutions have quasi-moduli parameters 
corresponding to the relative distance and phase 
between the constituent fractional instantons.
By summing over the Kaluza-Klein modes
of the quantum fluctuations around the bion backgrounds, 
we find that the effective action for
the quasi-moduli parameters is renormalized 
and becomes a function of the dynamical scale 
(or the renormalized coupling constant). 
Based on the renormalized effective action, 
we obtain the semiclassical bion contribution 
in a weak coupling limit 
by making use of the Lefschetz thimble method. 
We find that the non-perturbative contribution 
vanishes in the supersymmetric case and 
it has an imaginary ambiguity which is consistent with 
the expected infrared renormalon ambiguity 
in non-supersymmetric cases. 
This is the first explicit result indicating the relation 
between the complex bion and the infrared renormalon.
\end{abstract}

\maketitle

\newpage
\begin{spacing}{1.3}
\tableofcontents
\end{spacing}
\newpage 


\section{Introduction}
\label{sec:intro}
Understanding the non-perturbative aspects of 
quantum field theory (QFT) is 
one of the most long-standing problems in theoretical physics. 
Although lattice simulations may uncover 
non-perturbative phenomena 
such as confinement and dynamical mass generation 
in asymptotically free field theories 
including quantum chromodynamics (QCD), 
a systematic and analytical method to study
non-perturbative aspects of QFT 
has not yet been established in general.

The ``infrared renormalon" observed 
in the perturbative expansion in QFT 
\cite{'tHooft:1977am,Beneke:1998ui} 
is believed to be related to non-perturbative phenomena. 
In QCD, a specific set of Feynman diagrams 
with an internal chain of loops gives 
a factorial divergence of perturbation series 
with respect to the coupling constant $\alpha_{s}(\mu)$
renormalized at the energy scale $\mu$. 
The Borel transform of such a divergent series has 
singularities on the positive real axis of the Borel plane, 
leading to imaginary ambiguities of the Borel resummation. 
The first Borel singularity, 
which gives the leading imaginary ambiguity 
for small $\alpha_s(\mu)$, 
is located at $t = -2S_{I} / \beta_{0}$, 
and hence the corresponding imaginary ambiguity 
is proportional to 
$e^{2S_{I}/\beta_{0}} \approx |\Lambda_{\rm QCD}/\mu|^{4}$, 
where $\beta_{0} \ (<0)$ is the beta function coefficient, 
$S_{I}$ is the instanton action and 
$\Lambda_{\rm QCD}$ is the dynamical QCD scale. 
This indicates that the ambiguity arising 
in the perturbation series is associated 
with the low-energy non-perturbative physics. 
It is notable that the location of the Borel singularity is 
not twice of the instanton action $2S_{I}$, but $-2S_{I}/\beta_{0}$. Thus this singularity cannot be identified 
with an instanton--antiinstanton contribution 
unlike the quantum mechanical systems such as 
the double-well and sine-Gordon models \cite{Brezin:1977ab, 
Lipatov:1977cd, Bogomolny:1980ur,ZinnJustin:1981dx,Voros1,
Alvarez1,Alvarez2,Alvarez3,ZinnJustin:2004ib,ZinnJustin:2004cg,
Jentschura:2010zza,Jentschura:2011zza,Dunne:2013ada, Basar:2013eka,
Dunne:2014bca,Escobar-Ruiz:2015nsa,Escobar-Ruiz:2015nsa2,
Misumi:2015dua,Behtash:2015zha,Behtash:2015loa,Gahramanov:2015yxk,
Dunne:2016qix,Fujimori:2016ljw,Sulejmanpasic:2016fwr,Dunne:2016jsr,
Kozcaz:2016wvy,Serone:2016qog,Basar:2017hpr,Fujimori:2017oab,
Serone:2017nmd, Behtash:2017rqj,Costin:2017ziv,Fujimori:2017osz,
Alvarez:2017sza, Behtash:2018voa, Hatsuda:2017dwx}. 
The ambiguities of perturbation series associated with 
this type of Borel singularity in asymptotic-free QFTs 
are called ``infrared renormalon ambiguities".

The resurgence theory \cite{Ec1,Pham1,BH1,
Howls1,DH1,Costin1,Sauzin1,Sauzin2, Balian:1978ab,  DDP1, 
CNP1, DLS1, DP1, Takei1, CDK1, Takei2, Getm1, AKT1, Schafke1, Getm2}, which was originally discussed 
in the study of ordinary differential equations, 
has been investigated in various contexts
including matrix models and supersymmetric gauge theories
\cite{Marino:2006hs,Marino:2007te, 
Marino:2008ya, Marino:2008vx, Pasquetti:2009jg,Garoufalidis:2010ya, 
Drukker:2010nc, Aniceto:2011nu, Marino:2012zq,Schiappa:2013opa,
Hatsuda:2013oxa,Aniceto:2013fka,Santamaria:2013rua,Grassi:2014cla,
Couso-Santamaria:2014iia,Grassi:2014uua,Couso-Santamaria:2015wga, 
Aniceto:2015rua,Dorigoni:2015dha,Hatsuda:2015qzx,Franco:2015rnr,
Couso-Santamaria:2016vcc,Kuroki:2016ucm,Couso-Santamaria:2016vwq,
Arutyunov:2016etw, Aniceto:2018bis, Aniceto:2014hoa, Gukov:2016tnp, Honda:2016mvg, Honda:2016vmv, 
Gukov:2017kmk,Honda:2017qdb, Honda:2017cnz, Fujimori:2018nvz}.  
From the viewpoint of the resurgence theory, 
it has been conjectured in four-dimensional (4D) QCD(adj.) 
and two-dimensional (2D) ${\mathbb C}P^{N-1}$ models
compactified on $S^1$ with a small compactification radius
\cite{Argyres:2012vv, Argyres:2012ka, Dunne:2012ae,
Dunne:2012zk} 
that the renormalon could be identified as 
an object called the bion, 
which is composed of a pair of 
fractional instanton and anti-instanton. 
In these models, fractional instantons 
with fractional topological charges ($Q=1/N$) emerge 
\cite{Eto:2004rz,Eto:2006mz,Eto:2006pg,Bruckmann:2007zh,Brendel:2009mp,Bruckmann:2018rra} 
due to the $\mathbb Z_{N}$-symmetric 
Polyakov-loop holonomy in the compactified dimension 
(equivalent to the $\mathbb Z_{N}$-symmetric 
twisted boundary condition). 
The conjecture states that 
the Borel singularity corresponding 
to the bion could become the renormalon singularity at 
$-2S_{I}/\beta_{0}$ due to renormalization or decompactification. 
In spite of the recent progress on 
compactified ${\mathbb C}P^{N-1}$ models
with twisted boundary conditions
\cite{Dunne:2015eoa,Dunne:2015eaa,Dunne:2016nmc,Yamazaki:2017ulc,Evslin:2018yfm}
and the intensive studies on the resurgence in the 2D models \cite{Cherman:2013yfa,Cherman:2014ofa,Misumi:2014jua,Misumi:2014bsa,Misumi:2014rsa,Nitta:2014vpa,Nitta:2015tua,Behtash:2015kna,Dunne:2015ywa,Buividovich:2015oju,Misumi:2016fno, Demulder:2016mja,Sulejmanpasic:2016llc,Dorigoni:2017smz,Okuyama:2018clk}, 
the conjecture has not yet been verified even in the 2D sigma models.

In the ${\mathbb C}P^{N-1}$ quantum mechanics 
corresponding to the small compactification radius limit of
the ${\mathbb C}P^{N-1}$ model on ${\mathbb R} \times S^1$ 
with the twisted boundary condition, 
it was shown that the semiclassical contributions 
from the bion saddle points of the complexified action 
cancel the imaginary ambiguity 
in the Borel resummation of the perturbation series. 
Furthermore, it was confirmed that 
the full resurgent trans-series 
composed of the contributions 
from an infinite tower of the complex saddle points 
correctly gives the exact result 
\cite{Fujimori:2016ljw, Fujimori:2017oab, Fujimori:2017osz}. 
It is notable that the single complex bion solution is 
a composite of a kink and an anti-kink 
corresponding to the fractional instanton and 
fractional anti-instanton in 2D, respectively.  
This fact indicates that in the 2D $\C P^{N-1}$ sigma model, 
bion solutions composed of 
fractional instantons with opposite topological charges
are the saddle points 
corresponding to the perturbative Borel singularity. 
The question is whether the 2D complex bion can be 
identified as the infrared renormalon.

In this work, we discuss bions
in the ${\mathbb C} P^{N-1}$ sigma model on 
${\mathbb R} \times S^{1}$ with a twisted boundary condition. 
In particular, we show that 
the semiclassical bion contribution have the imaginary ambiguity,
which is consistent with that of the infrared renormalon. 
We start with the $\mathcal N =(2,2)$ supersymmetric (SUSY) model 
with a SUSY breaking deformation parameter $\delta \epsilon$. 
Generalization to non-supersymmetric cases is also 
discussed by introducing $n_{\rm F}$ copies of 
the fermionic degrees of freedom.  
We derive the real and complex bion solutions
by solving the complexified equations of motion
derived from the holomorphic action. 
The summation over the Kaluza-Klein (KK) modes 
of the quantum fluctuations around the bion configurations
correctly renormalizes the coupling constant 
in the bion effective action and gives 
the dimensionally transmuted dynamical mass scale 
$\Lambda_{{\mathbb C}P^{N-1}}$. 
Based on this renormalized effective action, 
we compute the semiclassical bion contributions 
to the vacuum energy. 
In the undeformed case ($\delta \epsilon = 0$), 
we find that the bion contribution vanishes 
due to a cancellation between the real and complex bions
when the fermion number $n_{\rm F}$ satisfies 
the condition $1 + N(n_{\rm F}-1)/2 \in \Z$.
This shows that the complex saddle point solutions 
play an important role to ensure $E=0$ 
in the vacuum of the supersymmetric model ($n_{\rm F}=1$). 
When the above condition for $n_{\rm F}$ is not satisfied, 
there are non-vanishing contributions to the vacuum energy 
with imaginary ambiguities.  
Even when the condition for $n_{\rm F}$ is satisfied, 
non-trivial bion contributions with the imaginary ambiguities
appear once the deformation parameter 
$\delta \epsilon$ is turned on.
These imaginary ambiguities are
in agreement with the expected renormalon ambiguity 
of the Borel resummation of the perturbation series. 
This implies that the bion solutions can be 
identified as the infrared renormalon in the 2D field theory. 

This paper is constructed as follows:
In Sec.\,\ref{sec:CP1}, 
the complex bion solutions in the ${\mathbb C}P^{1}$ model 
on ${\mathbb R}\times S^{1}$ are derived.  
In Sec.\,\ref{sec:bion}, the renormalized effective action on 
the quasi-moduli space of the bion solution is obtained. 
Based on the effective action, 
we calculate the contribution of the bions and 
compare it with the renormalon imaginary ambiguity. 
In Sec.\,\ref{sec:CPN}, 
the calculation for the ${\mathbb C}P^{1}$ model  
is extended to the ${\mathbb C}P^{N-1}$ models. 
Sec.\,\ref{sec:summary} is devoted to 
a summary and discussion.
In Appendix.\,\ref{appendix:QMS}, 
we illustrate the concept of the quasi-moduli space 
and valley solution using a simple zero-dimensional model.
In Appendices\,\ref{appendix:cp1_det} and 
\ref{appendix:cpn_oneloop},  
the detailed calculations of one-loop determinants 
are summarized for the $\C P^1$ and $\C P^{N-1}$ models, respectively. 
The Lefschetz thimble integral is summarized 
in Appendix \ref{app:LT}.


\section{$\C P^1$ sigma model and bion solutions}
\label{sec:CP1}

In the present and next sections, 
we investigate bions 
in the 2D $\C P^1$ sigma model on $\R \times S^1$
with emphasis on their relevance 
to the renormalon problem. 
We derive their semiclassical contributions 
in both supersymmetric and non-supersymmetric cases. 
The procedure here will be 
generalized to the $\C P^{N-1}$ model in Sec.\,\ref{sec:CPN}.


\subsection{$\C P^1$ sigma model on $\R \times S^1$}
Let us consider the 2D $\C P^1$ sigma model on $\R \times S^1$. 
For convenience, we start with 
the 2D $\mathcal N =(2,2)$ supersymmetric model.
The discussion in this section can also be generalized to 
the non-supersymmetric models 
with $n_{\rm F}$ copies of fermions\footnote{
When the models with $n_{\rm F}$ copies of fermions 
are reduced to quantum mechanics by compactification, 
they are called a quasi-exactly-solvable models
\cite{Kozcaz:2016wvy, Fujimori:2017osz}  
and enjoy a number of similar properties 
as the supersymmetric model, 
even though they are not supersymmetric. }. 
The bosonic degree of freedom $\varphi$ 
(the inhomogeneous coordinate of the target space $\C P^1$)
and the fermionic degrees of freedom $(\psi_l, \psi_r)$ 
form a chiral multiplet of 
the 2D $\mathcal N =(2,2)$ supersymmetry. 
The Lagrangian takes the form
\beq
\mathcal L \ = \ \frac{2}{g^2} \left[ G \left( \p \varphi \, 
\bar \p \bar \varphi + \bar \p \varphi  \, \p \bar \varphi 
- \bar \psi_l \D \psi_l - \bar \psi_r \bar \D \psi_r \right) 
+ \frac{1}{(1+|\varphi|^2)^4} \psi_l \bar \psi_l \psi_r \bar \psi_r \right]  
+ \mathcal L_{\rm top}, 
\label{eq:CP1Lag}
\eeq
where $\p$ and $\bar \p$ are the derivatives with respect to 
the Euclidean spacetime coordinates $z=x+iy$ and $\bar z = x- i y$
\beq
\p \equiv \frac{1}{2} (\p_x - i \p_y), \hs{10}
\bar \p \equiv \frac{1}{2} (\p_x + i \p_y),
\eeq
and $\D$ and $\bar \D$ are 
the pullbacks of the covariant derivatives 
onto 2D spacetime
\beq
\D \psi_l \equiv \left( \p + \Gamma \p \varphi  \right) \psi_l, \hs{10}
\bar \D \psi_r \equiv \left( \bar \p + \Gamma \bar \p  \varphi \right) \psi_r. 
\eeq
Here $\Gamma$ denotes the Christoffel symbol 
for the $\C P^1$ Fubini-Study metric $G$  
\beq
\Gamma \equiv G^{-1} \frac{\p}{\p \varphi} G 
= - \frac{2 \bar \varphi }{1+|\varphi|^2}, \hs{10}
G \equiv \frac{1}{(1+|\varphi|^2)^2}.
\eeq
The topological $\theta$-term $\mathcal L_{\rm top}$ is given by
\begin{equation}
\mathcal L_{\rm top} \ = \ \frac{i\theta}{\pi} G 
(\partial\varphi\bar\partial\bar\varphi 
-\bar\partial\varphi\partial\bar\varphi ).
\label{eq:topo_charge}
\end{equation}
For the moment, 
the coupling constant is denoted as $g^{2}$ in (\ref{eq:CP1Lag})
and we will regard it as the renormalized coupling $g_{R}^{2}$
when we discuss the renormalized bion effective action 
in Sec.\,\ref{subsec:eff_action}. 

This model admits 1/2 Bogomol'nyi-Prasad-Sommerfield (BPS) instanton solutions satisfying the BPS equation $\bar \p \varphi = 0$ 
\cite{Polyakov:1975yp}. 
They are characterized by non-trivial values of 
the topological charge. 
For an instanton solution 
with topological charge $k$, 
the Euclidean action is given by 
\beq
S \, \big|_{\mbox{\footnotesize{$k$-instanton}}} =  2\pi k i \tau, \hs{10}
\tau \equiv \frac{\theta}{2\pi} - \frac{i}{g^2}. 
\eeq

Throughout this paper, 
we regard $x$ and $y$ as the (non-compact) Euclidean time
and the (compact) spatial coordinate, respectively. 
Since the spatial direction $y$ is compactified on $S^1$ with 
the radius $R$, we must specify boundary conditions for the fields. 
Here we impose the following common twisted boundary condition 
for all the fields
\beq
\varphi(y+2\pi R) = e^{2 \pi i m R} \varphi(y), \hs{10}
\psi_{l,r}(y+2\pi R) = e^{2\pi i m R} \psi_{l,r}(y),
\label{eq:twisted_bc}
\eeq
where $R$ is the radius of $S^1$ and 
$m$ is the twist angle (in unit of $2\pi$) 
satisfying $0 < mR < 1$. 
It is well known that this twisted boundary condition works as 
the nontrivial holonomy for the global symmetry
in the compactified direction \cite{Eto:2004rz,Eto:2006mz,Eto:2006pg,
Dunne:2012ae}. 
We can find the potential of the 2D $\C P^1$ model 
with the twisted boundary condition 
by evaluating the action for the lightest mode 
in the KK expansion 
$\varphi(x,y)=\varphi_0 \hspace{1pt} e^{imy}$ 
with constant $\varphi_0$ 
\begin{equation}
V=\frac{m^2}{g^2}\frac{|\varphi_0|^2}{(1+|\varphi_0|^2)^2}.
\label{eq:mass_term}
\end{equation}
This potential exhibits two degenerate discrete minima 
at $\varphi_0=0$ (north pole) 
and $\varphi_0=\infty$ (south pole), 
in contrast to the vacuum manifold $\C P^1$ of 
the untwisted model. 

In the presence of the nontrivial background holonomy, 
the BPS instanton solution decomposes into 
fractional instantons. 
Each fractional instanton also satisfies the BPS equation 
and its explicit form is given by
\cite{Eto:2004rz,Eto:2006mz,
Eto:2006pg,Bruckmann:2007zh,Brendel:2009mp,Bruckmann:2018rra} 
\beq
\varphi_k = a \hspace{1pt} e^{m (x+iy)}, \hs{10} a \equiv  e^{-m x_0 + i \phi},
\eeq
where $a$ is a complex moduli parameter corresponding to 
the position $ x_0 = -(\log |a|)/m$ and internal phase 
$\phi = \arg a$ of the fractional instanton.
We note that this fractional instanton can be viewed as
a BPS kink solution connecting the two discrete vacua. 
Since the topological charge of a fractional instanton is 
smaller than that of an ordinary integer instanton, 
it may give a leading order non-perturbative contribution
to some physical quantities. 
The contribution of a single fractional instanton is of order 
\beq
\left| \exp \left( -S_{\rm k} \right) \right| = \exp \left( - \frac{2\pi mR}{g^2} \right),
\eeq
since the action of the fractional instanton 
is ${\rm Re} \, S_k = 2 \pi mR /g^2$.
However, such a non-perturbative contribution cannot be seen
in physical quantities such as the vacuum energy. 
This is because the path integral 
for the partition function 
in the zero temperature limit\footnote{
The zero temperature limit corresponds to the 
limit $\beta \rightarrow \infty$, 
where $\beta$ is the period of the Euclidean time $x$.}, 
from which the vacuum energy can be obtained, 
receives contributions only from configurations approaching 
the same field value at $x \rightarrow \pm \infty$. 
Therefore, the lowest order non-perturbative effect is given by 
a fractional instanton-antiinstanton pair. 
Such a composite configuration 
is called the bion \cite{Unsal:2007vu,Unsal:2007jx,
Shifman:2008ja,Poppitz:2009uq,Anber:2011de,Poppitz:2012sw,Misumi:2014raa}. 
It is notable that the bion configurations have 
no topological charge $(S_{\rm top} = 0)$, 
and tend to decay into the vacuum configuration. 
However, it becomes an approximate solution of 
the equation of motion 
when the constituent fractional instanton and 
anti-instanton are well separated. 
In the next subsection, 
instead of dealing with such an approximate solution of bion, 
we introduce a deformation parameter 
so that the equation of motion admits an exact bion solution. 


\subsection{A deformation and exact single bion solution}

To analyze the bion configuration and 
its contribution to the path integral, 
it is convenient to consider the following operator 
which is proportional to the height function
\beq
 \Delta \mu \ = \ m \frac{1-|\varphi|^2}{1+|\varphi|^2},
\label{eq:deformation}
\eeq
where $\Delta$ is the Laplacian on $\C P^1$ and 
$\mu \equiv m|\varphi|^2/(1+|\varphi|^2)$ is 
the moment map for the $U(1)$ symmetry
$\varphi \rightarrow e^{i \alpha} \varphi$. 
To calculate the generating function for $\Delta \mu$, 
we introduce the source term 
\beq
\delta \mathcal L \, = \, - \frac{\delta \epsilon}{2\pi R} \, 
\Delta \mu, 
\eeq
and evaluate the path integral for the partition function  
in the presence of $\delta \mathcal L$
\beq
Z(\delta \epsilon) = \int \mathcal D \varphi \, \exp \left( - S \right), \hs{10}
S \equiv \int d^2 x \left( \mathcal L + \delta \mathcal L \right). 
\label{eq:functionZ}
\eeq
We compactify the Euclidean time direction $x \sim x+\beta$ 
with the periodic boundary condition, 
and take a decompactification limit $\beta \to \infty$.
The vacuum expectation value 
$\langle \Delta \mu \rangle$ can be obtained 
by differentiating $Z(\delta \epsilon)$ with respect to 
the parameter $\delta \epsilon$ 
\beq
\langle \Delta \mu \rangle \ = \ \lim_{\delta \epsilon \rightarrow 0} \lim_{\beta \rightarrow \infty }  \frac{1}{\beta} \frac{\p}{\p \delta \epsilon} \log Z(\delta \epsilon),
\eeq
where 
$\langle \mathcal O \rangle$ denotes 
the expectation value of an operator $\mathcal O$ 
evaluated in one of the two vacua (corresponding to the classical vacuum $\varphi=0$).  
Since $\Delta \mu$ is not invariant 
under the SUSY transformation, 
the addition of $\delta \mathcal L$ can be regarded as 
a SUSY breaking deformation of the Lagrangian\footnote{
This deformation is motivated by the form of the potential 
induced when the Hilbert space is projected to the fermion number eigenspace in $\C P^1$ supersymmetric quantum mechanics \cite{Fujimori:2017osz}.}. 
In the deformed model, the vacuum energy can be expanded 
around the SUSY point $\delta \epsilon = 0$ as
\beq
E (\delta \epsilon) \ = \ - \lim_{\beta \rightarrow \infty} \frac{1}{\beta} \log Z(\delta \epsilon) \ = \ E^{(0)} + E^{(1)} \delta \epsilon + E^{(2)} \delta \epsilon^2 + \cdots,
\eeq
where the expansion coefficients are given by
\beq
E^{(0)} = 0 , \hs{5} 
E^{(1)} = - \langle \Delta \mu \rangle, \hs{5} 
E^{(2)} = - \frac{1}{2\pi R} \int d^2 x \, \Big[ \langle \Delta \mu(x) \Delta \mu(0) \rangle - \langle \Delta \mu \rangle^2 \Big], \hs{5} \cdots.
\eeq
The zeroth order expansion coefficient $E^{(0)}$ vanishes 
due to the unbroken supersymmetry at $\delta \epsilon = 0$. 
The first order expansion coefficient $E^{(1)}$ is given by
the vacuum expectation value $\langle \Delta \mu \rangle$,
which implies that $\langle \Delta \mu \rangle$ can also 
be interpreted as the response of the vacuum energy 
to the small SUSY breaking deformation.  
In general, the $n$-th order coefficients correspond to 
the $n$-point functions of $\Delta \mu$ 
integrated over the spacetime coordinates. 

To calculate $Z(\delta \epsilon)$, 
let us find saddle point solutions of the deformed action $S$ 
by solving the Euclidean equation of motion.  
The deformation term (\ref{eq:deformation}) 
causes a splitting of the two degenerate vacua 
of the undeformed model in such a way that 
only $\varphi = 0$ remains the global minimum of the potential. 
This implies that only saddle point solutions 
satisfying the boundary condition 
$\varphi \rightarrow 0$ at the spatial infinity 
$x \rightarrow \pm \infty$
can contribute to the partition function 
in the limit $\beta \rightarrow \infty$. 
The simplest solution which satisfies this boundary condition 
is a single bion configuration, 
whose explicit form is given by 
\beq
\varphi \, = \, \frac{\omega}{\sqrt{\omega^2-m^2}} 
\frac{e^{i m y + i \phi_0}}{\sinh \omega(x-x_0)},
\label{eq:bion_sol}
\eeq
where arbitrary constants $x_0$ and $\phi_0$ are 
moduli parameters corresponding to 
the overall position and phase and 
$\omega$ is the mass of the scalar field fluctuation around $\varphi=0$ 
\begin{equation}
\omega \, \equiv \, m \sqrt{1 + \frac{g^2 \delta\epsilon}{\pi m R}}. 
\label{eq:omega_cp1}
\end{equation}
This bion solution can be viewed as 
a bound state of fractional instantons 
with opposite topological charges. 
We can see this more clearly by rewriting the solution as
\beq
\varphi ~=~ e^{i m y} \left( e^{-\omega(x-x^{\rm rb}_-)-i\phi^{\rm rb}_-} 
+ e^{\omega(x-x^{\rm rb}_+)-i\phi^{\rm rb}_+} \right)^{-1} ,
\label{eq:kink_antikink_0}
\eeq
where the position and phase of 
the fractional instanton $(x^{\rm rb}_+, \phi^{\rm rb}_+)$ and 
those of the fractional anti-instanton 
$(x^{\rm rb}_-, \phi^{\rm rb}_-)$ are given by
\beq
x^{\rm rb}_\pm = x_0 \pm \frac{1}{2\omega} 
\log \frac{4\omega^2 }{\omega^{2}-m^{2}}, 
\hs{10}
e^{i \phi^{\rm rb}_\pm} = \pm e^{i\phi_0}.
\label{eq:bion_saddle}
\eeq
The superscript ``${\rm rb}$" indicates that 
these are the values of the parameters 
corresponding to the ``real bion".
In the weak coupling limit, 
which we will consider in the subsequent sections, 
the relative distance $|x^{\rm rb}_+-x^{\rm rb}_-|$ becomes large 
and diverges as $|x^{\rm rb}_+-x^{\rm rb}_-| \sim 
\frac{1}{m} \log \frac{1}{g^2}$. 
In such a situation, we can see that 
the bion solution is approximately given by 
a superposition of fractional instanton and anti-instanton
\beq
\varphi ~ \sim ~
\left\{ 
\begin{array}{ll} 
e^{\ m(z-x^{\rm rb}_-)+i\phi^{\rm rb}_-} & \mbox{for $x \approx x^{\rm rb}_- $} \\
\infty & \mbox{for $x \approx x_0$} \\
e^{-m(\bar z-x^{\rm rb}_+)+i\phi^{\rm rb}_+} & \mbox{for $x \approx x^{\rm rb}_+$} 
\end{array} \right..
\label{eq:superposition}
\eeq

In the semiclassical method, we need to take into 
account of all possible saddle point solutions 
not only in the original configuration space 
but also in the complexified field space.  
The $\C P^1$ sigma model can be complexified 
by regarding $(\varphi, \bar \varphi)$ as 
independent holomorphic coordinates 
$(\varphi,\tilde \varphi)$ of 
the complexified target space 
$(\C P^1)^\C \cong SU(2)^\C/U(1)^\C \cong  SL(2,{\C})/\C^{*}
\cong T^* \C P^1$.
Then we can obtain another single bion solution
\cite{Fujimori:2016ljw}
\beq
\varphi = \frac{\omega}{\sqrt{\omega^2-m^2}} 
\frac{i e^{i m y + i \phi_0}}{\cosh \omega(x-x_0)},
\hs{10}
\tilde{\varphi} = \frac{\omega}{\sqrt{\omega^2-m^2}} 
\frac{i e^{i m y - i \phi_0}}{\cosh \omega(x-x_0)}.
\label{eq:complex_bion_sol}
\eeq
This is not a proper solution before the complexification 
since $\tilde \varphi$ is not the complex conjugate of $\varphi$, 
so that the saddle point solution \eqref{eq:complex_bion_sol} 
is called ``the complex bion",
whereas \eqref{eq:bion_sol} is called ``the real bion".
The complex bion solution can also be rewritten 
into the same form as the real bion in 
Eq.\,\eqref{eq:kink_antikink_0}
but with the following complex values of
the relative separation and phase 
\beq
x^{\rm cb}_\pm = x_0 \pm \frac{1}{2\omega} 
\left(\log \frac{4\omega^2}{\omega^{2}-m^{2}} + i \pi \right), 
\hs{10}
e^{i \phi^{\rm cb}_\pm} = e^{i\phi_0}.
\label{eq:complex_bion_saddle}
\eeq 
The superscript ``${\rm cb}$" stands for the ``complex bion".
In the next section, we will derive the semiclassical contribution 
from these bion solutions by calculating the associated one-loop 
determinants and quasi-moduli integrals.


\section{Non-perturbative bion contribution to partition function}
\label{sec:bion}
In this section, 
we calculate the non-perturbative contributions 
of the real and complex bions. 
We focus only on the leading non-perturbative contributions 
in the weak coupling limit, 
and hence we always ignore irrelevant terms 
in the limit $g \rightarrow 0$ in the following.  

\subsection{Quasi-moduli space of single bion configuration}

To calculate the bion contributions to the partition function, 
we need to complexify the configuration space 
and evaluate the path integral 
along an appropriate path integral contour 
emanating from each bion saddle point. 
Although, in principle, such a contour can be determined 
by the Lefschetz thimble method, 
it is not easy to apply it 
in the infinite dimensional configuration space. 
Instead, let us consider 
a reduction of the degrees of freedom 
from the infinite dimensional field space to 
a finite dimensional subspace called ``the quasi-moduli space."

In the weak coupling limit $g \rightarrow 0$, 
almost all massive modes can be integrated out 
and their contribution can eventually 
be expressed as one-loop determinants. 
However, there are four modes 
which become massless in the limit $g \to 0$. 
Two of them are the exact zero modes 
associated with the two moduli parameters: 
the overall position and phase $(x_0, \phi_0)$. 
The others are called the quasi zero modes,  
corresponding to the relative position and phase of 
the constituent fractional instantons.  
To evaluate the integral along such ``nearly flat directions" 
(flat directions in the limit of $g \to 0$),  
let us define ``valley solution" $\varphi_{B}(\eta)$ 
(quasi-solution) \cite{Aoyama:1991ca,Aoyama:1994sk,Aoyama:1995ca,Aoyama:1997qk,Aoyama:1998nt} 
as a bion ansatz satisfying the following properties:
\begin{itemize}
\item
$\varphi_B(\eta)$ is parameterized 
by the positions and phases 
of the constituent fractional instantons 
$\eta^\alpha = (x_-,\, \phi_-,\, x_+,\, \phi_+)$, 
\item $\varphi_B(\eta)$ becomes the exact real and complex bion solutions
when these quasi-moduli parameters $\eta^\alpha$ are 
at the saddle point values (\ref{eq:bion_saddle}) and 
(\ref{eq:complex_bion_saddle}). 
\item
$\varphi_B(\eta)$ satisfies the equation of motion 
up to a linear combination of $\p \varphi_B/\p \eta^\alpha$. 
In other words, it is a solution of the equation 
\beq
\frac{\delta S}{\delta \varphi} \, \bigg|_{\varphi = \varphi_B} \!
= \, G \overline{A^\alpha \frac{\p \varphi_B}{\p \eta^\alpha}}.
\label{eq:valley}
\eeq
Here, the metric $G$ is the ${\mathbb C}P^1$ Fubini-Study metric, 
and the coefficients of the linear combination 
$A^\alpha$ can be determined by taking the inner product of 
Eq.\,\eqref{eq:valley} and $\p_\alpha \varphi_B$ 
\beq
A^\alpha = \overline{\frac{\p S_{\rm eff}}{\p \eta^\beta}} \, 
g^{\bar \beta \alpha},
\label{eq:coef_valley}
\eeq
where $S_{\rm eff}(\eta)$ is the bion effective action 
and $g_{\alpha \bar \beta}$ is the induced metric  
\beq
S_{\rm eff}(\eta) = S[\varphi_B], \hs{10} 
g_{\alpha \bar \beta} = \int d^2 x \, G \hspace{1pt} \frac{\p \varphi_B}
{\p \eta^\alpha} \overline{\frac{\p \varphi_B}{\p \eta^\beta}}. 
\eeq
\end{itemize}
We call the set of valley solutions 
``the quasi-moduli space $\mathcal M$". 
The quasi-moduli parameters $\eta^\alpha$ 
can be regarded as coordinates of $\mathcal M$
and $g_{\alpha \bar \beta}$ is the metric on $\mathcal M$.
Roughly speaking, the quasi-moduli space 
is a valley of the action, 
where the gradient of S is tangent to the valley. 
In Appendix \ref{appendix:QMS}, 
the concept of the quasi-moduli space is 
explained in more detail 
by using an example of a simple zero dimensional model.

To evaluate the path integral, 
let us decompose the bosonic degree of freedom 
into the bion background $\varphi_B(\eta)$ 
(parametrized by the quasi-moduli parameters $\eta^\alpha$)
and a fluctuation field $\delta \varphi$ 
which is orthogonal to the quasi-zero modes  
\beq
\varphi = \varphi_{B}(\eta) + \delta \varphi, \hs{10} 
\int d^2 z \, G \overline{\frac{\p \varphi}{\p \eta^\alpha}} 
\delta \varphi = 0. 
\label{eq:orthogonal}
\eeq
Then the path integral decomposes into that for $\delta \varphi$
and the quasi-moduli integral over $\mathcal M$. 
Note that $\delta \varphi$ stands for all the modes 
which remain massive in the weak coupling limit 
$g \rightarrow 0$.
It is convenient to redefine 
the bosonic fluctuation\footnote{
It would be more convenient to use the Riemann normal coordinates 
\cite{AlvarezGaume:1981hn} or K\"ahler normal coordinates \cite{Higashijima:2000wz, Higashijima:2002fq} 
for higher loop computations
} and fermionic fields as 
\beq
\delta \varphi = g (1+|\varphi_B|^2) \, \xi, \hs{10}
\psi_{l,r} = g (1+|\varphi_B|^2) \, \chi_{l,r}. 
\eeq
Thanks to the definition of the valley solution \eqref{eq:valley}
and the orthogonality condition in \eqref{eq:orthogonal}, 
no linear term of the fluctuation fields appears 
in the action expanded in powers of $g$
\beq
S = S_{\rm eff}(\eta) - \int d^2 z \Big[ 
\mathcal L_{\rm fluc} (\xi, \bar \xi) 
+ 2 \bar \chi_l \nabla \chi_l + 2 \bar \chi_r \bar \nabla \chi_r \Big] 
+ \mathcal O(g^2) ,
\label{eq:S0}
\eeq
where $S_{\rm eff}(\eta)= S[\varphi_{B}(\eta)]$
is the bion effective action and  
$\mathcal L_{\rm fluc} (\xi, \bar \xi)$ 
is a quadratic term of the bosonic fluctuation, 
whose variation gives 
the following linearized equation of motion for the fluctuation 
\beq
\Delta_B \ba{c} \xi \\ \bar \xi \ea = 0, 
\eeq
with 
\beq
\Delta_B =
\ba{cc} 
\{ \nabla , \bar \nabla \} & 
-4\varphi_B \mathcal E \\
-4\overline{\varphi_B \mathcal E} & 
\{ \bar \nabla^\ast, \nabla^\ast \}
\ea 
+ \frac{1}{(1+|\varphi_B|^2)^2}
\ba{cc} 
\phantom{-} \p_i \varphi_B \p_i \bar \varphi_B & 
- \p_i \varphi_B \p_i \varphi_B \\ 
- \p_i \bar \varphi_B \p_i \bar \varphi_B & 
\phantom{-} \p_i \bar \varphi_B \p_i \varphi_B
\ea 
+ \mathcal O(g^2). 
\label{eq:DeltaB}
\eeq
Here we have defined $\mathcal E$ as\footnote{
$\mathcal E = 0$ is the equation of motion 
without the deformation term.}
\beq
\mathcal E = \frac{\D \bar \p \varphi_B}{1+|\varphi_B|^2} 
= \frac{1}{1+|\varphi_B|^2}\left( \p - \frac{2 \bar \varphi_B}
{1+|\varphi_B|^2} \p \varphi_B \right) \bar \p \varphi_B.
\label{eq:eom_0}
\eeq 
The differential operators $\nabla$, $\bar \nabla$, $\nabla^\ast$ 
and $\bar \nabla^\ast$ are given by
\beq
\nabla = \p + 2 i A_z , \hs{5}
\bar \nabla = \bar \p + 2 i A_{\bar z}, \hs{5}
\nabla^\ast = \p - 2 i A_z , \hs{5}
\bar \nabla^\ast = \bar \p - 2 i A_{\bar z},
\label{eq:covariant_derivative1}
\eeq
with $A_z=(A_x-iA_y)/2$ and $A_{\bar z} = (A_x+iA_y)/2$ 
defined as\footnote{
$A_i$ are auxiliary gauge fields 
in the gauged linear sigma model realization of 
the $\C P^1$ sigma model, 
and the corresponding field strength is the topological charge density.
}
\beq
A_i = \frac{i}{2} \frac{\bar \varphi_B \p_i \varphi_B - \varphi_B 
\p_i \bar \varphi_B}{1+|\varphi_B|^2} . 
\label{eq:covariant_derivative}
\eeq
It is notable that the expression (\ref{eq:DeltaB}) is 
valid for arbitrary values of $\delta\epsilon$ 
as long as we work in the weak coupling limit $g^{2}\to 0$ 
since all the $\delta\epsilon$ dependence is included 
in the $\mathcal O(g^2)$ term.

Let $S_{\rm q}$ be the quantum correction 
induced by the bosonic and fermionic fluctuations 
\beq
\exp \left( - S_{\rm q} \right) \ = \ \int 
\D \Phi \hspace{1pt} \D \Psi \, \exp \big[ \! - \left( S - S_{\rm eff} \right) \big],
\eeq
where $\D \Phi \hspace{1pt} \D \Psi$ 
denotes the path integral measure for the fluctuation. 
In principle, this can be evaluated 
by the standard perturbation expansion. 
In particular, 
the leading order term is given by the one-loop determinants
\beq
\exp \left( - S_{\rm q} \right) \ = \ \frac{\det \Delta_F}{\det' \!\Delta_B} \left[ 1 + \mathcal O(g^2) \right].
\eeq
$\det \Delta_F$ is the fermionic one-loop determinant
and $\det' \Delta_B$ is the bosonic one-loop determinant  
excluding the quasi-zero modes. 
Then the single bion contribution to the partition function 
can be rewritten as
\beq
Z_1 = \int_{\mathcal M} dv \, \exp \left( - S_{\rm eff} - S_{\rm q} \right). 
\eeq
where $dv$ denotes the volume form 
on the quasi-moduli space $\mathcal M$.


\subsection{Single bion effective action and renormalization}
\label{subsec:eff_action}
Let us first calculate 
the classical effective bion action 
$S_{\rm eff}(\eta) = S[\varphi_B(\eta)]$
in the weak coupling limit.
As we can see from Eqs.\,\eqref{eq:bion_saddle} 
and \eqref{eq:complex_bion_saddle}, 
the saddle points run away to infinity as $g \rightarrow 0$. 
To focus on the vicinity of the saddle points 
in the configuration space, 
we take the weak coupling limit as
\beq
g \rightarrow 0 ~~ \mbox{with fixed} ~~ 
\delta x_r \ \equiv \ x_+-x_- - 
\frac{1}{\omega} \log \frac{4 \omega^2}{\omega^2 - m^2},
\label{eq:deviation_from_saddle}
\eeq
where $\delta x_r$ is the deviation of the relative distance 
from the value at the real bion saddle point:
$\delta x_r = 0$ and $\delta x_r = \pi i/\omega$
for the real and complex bions, respectively   
(see Eqs.\,\eqref{eq:bion_saddle} and \eqref{eq:complex_bion_saddle}).  
Since the relative distance $|x_+-x_-|$ is always large 
in this limit, we can regard
\beq
\exp(-\omega |x_+-x_-| \, ) \sim \mathcal O(g^2).
\eeq

In the weak coupling limit, 
the valley equation \eqref{eq:valley}
can be solved as
\beq
\varphi_B(\eta) ~=~ e^{i m y}
\left( e^{-\omega(x-x_-)-i\phi_-} 
+ e^{\omega(x -x_+)-i\phi_+} \right)^{-1}. 
\label{eq:bion_ansatz}
\eeq
We can show that 
this satisfies the valley equation 
in the weak coupling limit
by substituting $\varphi_B(\eta)$ 
into the equation of motion $\delta S/\delta \varphi$ 
and by checking that it is of order $\mathcal O(g^2)$ 
everywhere on $\R \times S^1$. 
It is notable that 
this bion ansatz becomes the exact bion solution 
\eqref{eq:bion_sol} or \eqref{eq:complex_bion_sol} 
when the quasi-moduli parameters 
$\eta^\alpha =(x_\pm, \phi_\pm)$ 
sit at the saddle point \eqref{eq:bion_saddle} 
or at \eqref{eq:complex_bion_saddle}, respectively. 
As in the case of the saddle point configuration, 
$\varphi_B(\eta)$ can be viewed as a superposition of 
fractional instantons \eqref{eq:superposition}. 

By substituting the bion ansatz \eqref{eq:bion_ansatz} 
into the original action with the deformation term, 
we obtain the single bion effective action\footnote{\label{footnote:seff}
It is notable that  the effective action contains 
a subleading term of order $\mathcal O(1)$ 
even though it is derived from 
the leading order valley solution \eqref{eq:bion_ansatz}.
This is justified because  
the subleading correction $\delta \varphi_B$ does not 
contribute to the subleading term in $S_{\rm eff}$, 
thanks to the definition of the valley solution \eqref{eq:valley}
and the orthogonality between $\delta \varphi_B$ 
and the quasi-zero modes $\p \varphi_B / \p \eta^\alpha$.
} 
\beq
S_{\rm eff}(\eta) =  \frac{4\pi R}{g^2} 
\Big[ m - 2 m \cos (\phi_+-\phi_-) e^{-m(x_+-x_-)} 
\Big] + 2 \delta\epsilon m (x_+-x_-)
 + {\mathcal O} \left( g^2 \right). 
\label{eq:bare_eff}
\eeq
The first term in $[\cdots]$ 
is the asymptotic value of the classical bion action, 
the second term in $[\cdots]$ is the interaction between 
the constituent fractional instantons at large separation, 
and the term proportional to $(x_+-x_-)$
represents the confining potential 
due to the deformation potential. 
As mentioned above, 
we have neglected terms 
of order $\mathcal O(e^{-2m(x_+-x_-)}/g^{2})$
since the leading order contribution of the quasi-moduli integral 
comes from the vicinity of the saddle points, 
where $\mathcal O(e^{-2m(x_+-x_-)}/g^{2}) \sim \mathcal O(g^2)$. 

Next, let us integrate out 
the bosonic and fermionic fluctuations in Eq.\,\eqref{eq:S0}.
By evaluating the Gaussian integral for the fluctuations, 
the single bion contribution to the partition function 
can be rewritten into the quasi-moduli integral
\beq
Z_1 = \int_{\mathcal M} d^4 \eta \, \mathcal J \, 
\frac{\det \, \Delta_F}{\det' \Delta_B} 
\exp \left( -S_{\rm eff} \right) + \mathcal O(g^2), 
\eeq
where $\mathcal J$ is volume factor 
associated with the metric of the quasi-moduli space, 
$\det \Delta_F$ is the fermionic one-loop determinant, 
and $\det' \Delta_B$ is the bosonic one-loop determinant  
excluding the quasi-zero modes. 
The divergent integral with respect 
to the center of mass position $x_+ + x_-$ can be regularized 
by compactifying the $x$-direction as $x \sim x + \beta$. 
Then, the single bion contribution 
to the vacuum energy can be written as
\beq
E_1 \ \equiv \ - \lim_{\beta \rightarrow \infty} \frac{1}{\beta} \frac{Z_1}{Z_0} \ = \, -2 \pi \int_{\mathcal M_r} dx_r d \phi_r \hspace{1pt} 
\mathcal J \hspace{1pt} \exp \left( - S_R \right),
\label{eq:Z1Z2ratio}
\eeq
where $\mathcal M_r$ is 
the space of relative quasi-moduli parameterized by
$x_r \equiv x_+ - x_-$ and $\phi_r \equiv \phi_+ - \phi_-$. 
The quantum corrections from the fluctuations
is included into the renormalized effective action 
\beq
S_R(x_r,\phi_r) \ \equiv \ S_{\rm eff}(x_r,\phi_r) 
- \log \frac{\det \Delta_F}{\det \Delta_{F}^0}
+ \log \frac{\det' \! \Delta_B}{\det \Delta_{B}^0}, 
\label{eq:SRdef}
\eeq
where $\det \Delta_{F}^0$ and $\det \Delta_{B}^0$ are the 
one-loop determinants around the trivial saddle point $\varphi=0$. 

Now let us calculate the one-loop determinants. 
Since the background bion ansatz \eqref{eq:bion_ansatz}  
is independent of the compactified coordinate $y$ 
(except the twisting factor dictated 
by the twisted boundary condition), 
it is convenient to decompose 
the bosonic and fermionic fluctuations 
into infinite towers of KK modes. 
The total contribution from 
the non-zero bosonic KK modes takes the form
\beq
\log \frac{\det' \! \Delta_B}{\det \Delta_{B}^0} \bigg|_{\rm KK}
&=& 2 \sum_{n=1}^\infty \bigg[ X_n
+ Y_n \cos \phi_r \, e^{-mx_r} +  \mathcal O(g^2) \bigg],
\label{eq:logdKK}
\eeq
where $X_n$ and $Y_n$ are respectively calculated 
in Appendices \ref{app:cp1_LO} and \ref{app:cp1_LKK} as
\beq
X_n = \log \frac{\frac{n}{R}-m}{\frac{n}{R}+m} , \hs{10} 
Y_n = {4mR\over{n}} + \mathcal O(n^{-2}).
\eeq  
By using the zeta function regularization shown in (\ref{eq:logdKKapp}) and (\ref{eq:zeta}), 
we find that 
\beq
\sum_{n=1}^\infty X_n =  - 2 m R \, \log R \Lambda_0 
+ \log \frac{\Gamma (1+ m R)}{\Gamma(1-mR)}, \hs{10} 
\sum_{n=1}^\infty Y_n = 4 m R \log R \Lambda_{0} + \cdots,
\label{eq:logdKK2}
\eeq
where $\Lambda_{0}$ is a parameter 
corresponding to the cutoff scale 
and $\cdots$ denotes terms 
without $\Lambda_0$-dependence, 
which give only subleading contributions 
in the weak coupling limit. 
The $\Lambda_0$-dependent terms 
in the one-loop determinant 
can be absorbed into the bare coupling constant, 
{\it i.e.} 
they are canceled by appropriate counter terms. 
The renormalized effective action 
can be written in terms of the dynamical scale $\Lambda$ 
defined by
\beq
\Lambda = \Lambda_0 \, \exp \left( -\frac{\pi}{g_0^2} - \frac{i}{2} \theta_0 \right), \hs{10}
\left(\tau_0 = \frac{1}{2\pi i} \log \frac{\Lambda_0^2}{\Lambda^2} \right),
\eeq
where the symbols with subscript 0 denote the bare parameters.
In the following, we explicitly denote the coupling constant $g$ 
which has been used above 
as the renormalized coupling constant $g_R$ at the scale $1/R$
\beq
\frac{1}{g_R^2} = -\frac{1}{\pi} \log |R \Lambda| \hs{10} \left( \ = ~ \frac{1}{g_0^2} -\frac{1}{\pi} \log |R \Lambda_0| \right),
\eeq
and interpret that the UV divergent terms in the one-loop 
determinant are canceled by the corresponding counter terms 
in the renormalized perturbation theory.

The contributions of the fermionic KK modes 
with positive and negative KK momenta cancel out 
(see Appendix \ref{app:cp1_F} for details)
\beq
\log \frac{\det \Delta_F}{\det \Delta_{F}^0} \bigg|_{\rm KK} = 0.
\label{eq:DeltaFKK}
\eeq
The contributions of the bosonic and fermionic KK zero modes 
(and the volume factor $\mathcal J$) are essentially the same 
as in the 1D case \cite{Fujimori:2016ljw} 
\beq
\mathcal J \, \frac{\det' \! \Delta_B}{\det \Delta_{B}^0} 
\bigg|_{n=0} ~\approx~ \left( \frac{4m^2 R}{g_R^2} \right)^2 + \cdots,
\hs{10}
\frac{\det \Delta_F}{\det \Delta_{F}^0} ~\approx 
e^{- 2 m x_{r}} + \cdots, 
\label{eq:jacobian_cp1}
\eeq
where $\cdots$ denotes terms which are irrelevant in the weak 
coupling limit $g_{R}^{2} \rightarrow 0$. 

Finally, by combining the one-loop determinants 
\eqref{eq:logdKK}, \eqref{eq:DeltaFKK} and \eqref{eq:jacobian_cp1}, 
we obtain the renormalized effective bion action (\ref{eq:SRdef}), from which we find that 
the integrand for the quasi-moduli integral \eqref{eq:Z1Z2ratio} 
is given by
\beq
\mathcal J \exp\left(- S_R \right) \ = \ \left( \frac{4m^2 R}
{g_R^2} \right)^2 \exp \left( -X + \frac{8\pi m R}{g_R^2} 
\cos \phi_r \, e^{-mx_r} - 2 m \epsilon x_r \right) + \cdots,
\eeq
where $\epsilon$ is the effective deformation parameter, 
which is shifted due to the fermionic contribution
\beq
\epsilon \equiv 1 + \delta \epsilon,
\label{eq:eff_deformation}
\eeq
and $X$ is the constant term 
\beq
X &\equiv& \frac{4\pi m R}{g_R^2}
+ 2 \log \frac{\Gamma (1+mR)}{\Gamma(1-mR)}.
\eeq
This implies that the single bion contribution is of order
\beq
E_1 \ \propto \ |R \Lambda|^{4m R} \, = \ e^{-4mR\pi/g_R^2(R)}. 
\eeq
Here we denote $g_{R}^2$ as $g_{R}^{2}(R)$ to emphasize 
that it is renormalized at the energy scale $\sim 1/R$.
The renormalized effective action $S_R$ can be 
rewritten in terms of $\log|R\Lambda|$ 
by replacing $1/g_{R}^{2}$ with $-\frac{1}{\pi} \log |R\Lambda|$.
We note that the emergence of 
the renormalized coupling (or the dynamical scale $\Lambda$)
is a consequence of the renormalization procedure 
in the semiclassical calculation of bion contributions. 
{\it The renormalized coupling constant is NOT inserted by hand
but it naturally appears in the semiclassical calculation.}


\subsection{Lefschetz thimble analysis and imaginary ambiguity}
Next let us calculate the single bion contribution to the 
vacuum energy $E_1$ by evaluating the quasi-moduli integral
\beq
E_1 \, = \, 
-2 \pi \left( \frac{4 m^2 R}{g_R^2} \right)^2 e^{-X} \int d x_r 
d \phi_r \, \exp \left(  {8\pi mR\over{g_{R}^{2}}} \cos \phi_r \, 
e^{-m x_r} - 2 m \epsilon x_r \right) + \cdots,
\label{eq:thimble_integral_cp1}
\eeq
where $\cdots$ denotes irrelevant terms 
in the weak coupling limit.
Since this is essentially the same integral 
as in the case of the $\C P^1$ quantum mechanics,
we can apply the Lefschetz thimble method \cite{Witten:2010cx,Cristoforetti:2013wha,Fujii:2013sra,Tanizaki:2014tua,Tanizaki:2014xba,Kanazawa:2014qma,Tanizaki:2015tnk,DiRenzo:2015foa,Fukushima:2015qza,Tanizaki:2015rda,Fujii:2015bua,Alexandru:2016gsd,Tanizaki:2016xcu}
in the same manner as in the 1D case \cite{Fujimori:2016ljw}
to evaluate the quasi-moduli integral. 
As shown in Ref.\,\cite{Fujimori:2016ljw} and 
briefly summarized in Appendix~\ref{app:LT}, 
$S_R$ has saddle points corresponding to 
the real and complex bions 
and their contributions to $E_1$ 
contain an imaginary ambiguity 
due to the Stokes phenomena
\beq
E_1 \, = \,
- 2 m \frac{\Gamma(\epsilon)}{\Gamma(1-\epsilon)} \left[ 
\frac{4 \pi m R}{g_R^2} \frac{\Gamma(1-mR)}{\Gamma(1+mR)} \right]^2 
\left( {4\pi mR\over{g_{R}^{2}}} e^{\pm \frac{\pi i}{2}} \right)^{-2 \epsilon} 
|R \Lambda|^{4m R} + \cdots,
\label{eq:vacuum_energy_cp1}
\eeq
where the upper (lower) sign corresponds to the positive 
(negative) imaginary part given to $g_R^2$ 
in order to avoid the coupling constant being on the Stokes line. 
For the $\Z_2$-twisted boundary condition ($m=1/(2R)$),
the result is of order $|R \Lambda|^{2}$, 
which is the expected order of 
the well-known imaginary ambiguity 
from the infrared renormalon. 

The resurgence theory states that 
there is a cancellation between 
the imaginary part of the non-perturbative contribution
${\rm Im} \, E_{\rm n.p.}$ 
and that of the perturbation series 
${\rm Im} \, E_{\rm pert}$. 
Since the single bion imaginary part 
${\rm Im} \, E_1$ is the dominant term in 
${\rm Im} \, E_{\rm n.p.}$, 
it cancels the leading order part of ${\rm Im} \, E_{\rm pert}$. 
This implies that the difference of ${\rm Im} \, E_{\rm pert}$ 
for positive and negative ${\rm Im} \, g_R$ has
the following asymptotic behavior in the weak coupling limit
$g_R \rightarrow 0$
\beq
\Delta {\rm Im} \, E_{\rm pert} \ \sim \ 
- \Delta {\rm Im} \, E_1 
\ = \ -4\pi m \left[ \frac{1}{\Gamma(1-\epsilon)} 
\frac{\Gamma(1-mR)}{\Gamma(1+mR)} \right]^2 \left( \frac{4\pi m R}
{g_R^2} \right)^{2(1-\epsilon)}  |R\Lambda|^{4mR}.
\eeq
From this imaginary part, 
we can read off the large order behavior of 
the perturbation series as 
\beq
E_{\rm pert} \, = \ \sum_{n=0}^\infty a_n \left( \frac{g_R^2}
{4 \pi m R} \right)^n \hs{3} \mbox{with} \hs{3}
a_n \rightarrow - 2m \left[ \frac{1}{\Gamma(1-\epsilon)} 
\frac{\Gamma(1-mR)}{\Gamma(1+mR)} \right]^2 
\Gamma( n + 2(1 - \epsilon)). 
\label{eq:large_order}
\eeq
This large order behavior of 
the perturbation series is the prediction of 
the resurgence theory. 
In the limit $R \rightarrow 0$ with 
fixed $g_{\rm 1d}^2 \equiv  g_R^2/(2 \pi R)$, 
this system reduces to the $\C P^1$ quantum mechanics
and it has been shown that 
the 1D limit of Eq.\,\eqref{eq:large_order} 
is the correct large order behavior 
in quantum mechanics \cite{Fujimori:2017oab}.
It would be interesting to check 
if the result \eqref{eq:large_order} 
obtained by the resurgence argument 
gives the consistent large order behavior of 
the perturbation series also in the 2D case. 

Next, let us consider the generalization 
to the case with $n_{\rm F}$ copies of fermions. 
Since in this case, the fermionic contribution 
to the bion effective action is multiplied by $n_{\rm F}$, 
the single bion contribution $E_1$ can be 
obtained by redefining the constant $\epsilon$ 
in Eq.\,\eqref{eq:vacuum_energy_cp1} as
\beq
\epsilon \, \equiv \, 1 + \delta \epsilon \, \rightarrow \, 
\epsilon \, \equiv \, n_{\rm F} + \delta \epsilon. 
\eeq 
Expanding $E_1$ around 
$\epsilon = n_{\rm F} = 1, \, 2 , \,\cdots$, 
we find that 
\beq
E_1 \ = \ C |R \Lambda|^{4m R} \Bigg[ \, \delta \epsilon + 
2 \left( \psi(n_{\rm F}) - \log \frac{4\pi m R}{g_R^2} 
\mp \frac{\pi i}{2} \right) \delta \epsilon^2 + \cdots \Bigg], 
\label{eq:epsilon_expansion}
\eeq
where $\psi(n_{\rm F}) \equiv \p_\epsilon \log \Gamma(\epsilon) \, 
|_{\epsilon = n_{\rm F}} = -\gamma + \sum_{r=1}^{n_{\rm F}-1}
\frac{1}{r}$ is the digamma function and 
\beq
C \equiv - 2m \left[ \Gamma(n_{\rm F}) \frac{\Gamma(1-mR)}{\Gamma(1+mR)} 
\left(\frac{4\pi m R}{g_{R}^2}\right)^{1-n_{\rm F}} \right]^2 .
\eeq
Eq.\,\eqref{eq:epsilon_expansion} implies that 
the single bion contribution vanishes for $\delta \epsilon =0$.
In the supersymmetric case $n_{\rm F}=1$, this is consistent with 
the fact that the vacuum energy vanishes.  
The absence of the non-perturbative correction 
at $\epsilon = n_{\rm F} = 1, \, 2 , \,\cdots$ 
is due to the cancellation between 
the real and complex bion saddle points\footnote{
This phenomenon can also be seen 
in some quantum mechanical models
in which the so-called quasi-exact-solvability plays an important role 
\cite{Kozcaz:2016wvy, Fujimori:2017osz}.}.
This cancellation happens because of 
a particular high symmetry at these points, 
whereas the non-perturbative corrections exist 
ubiquitously anywhere away from these particular points. 
We can see from Eq.\,\eqref{eq:epsilon_expansion} 
that a non-trivial non-perturbative correction 
and an imaginary ambiguity appear 
in the leading and subleading order terms
in the small $\delta \epsilon$ expansion, respectively.
We emphasize that 
for the $\Z_{2}$-twisted boundary condition,
the imaginary ambiguity is of order $|R \Lambda|^{2}$, 
which is consistent with that from the infrared renormalon. 
This result indicates that the renormalon ambiguity 
can be canceled by the bion contribution, 
and hence the bion could be identified 
as the infrared renormalon.


\section{Generalization to $\C P^{N-1}$ model}
\label{sec:CPN}
In this section, we consider the generalization of 
the analysis in the previous section 
to the $\C P^{N-1}$ model.
We compute the single bion contribution to the vacuum energy 
by embedding the single bion solutions of the $\C P^1$ model, 
calculating the one-loop determinant 
and evaluating the quasi-moduli integral. 

\subsection{Embedding single bion solution}
Let us consider the 2D $\mathcal N =(2,2)$ 
$\C P^{N-1}$ sigma model 
described by the Lagrangian
\beq
\mathcal L &=& \frac{2}{g^2} \left[ G_{a \bar b} \left( \p \varphi^a 
\bar \p \bar \varphi^{\bar b} + \bar \p \varphi^a 
\p \bar \varphi^{\bar b} - \bar \psi_l^{\bar b} \, 
\D \psi_l^a - \bar \psi_r^{\bar b} \, \bar \D \psi_r^a \right) 
+ \frac{1}{2} R_{a \bar b c \bar d} \, \psi_l^a 
\bar \psi_l^{\bar b} \psi_r^c \bar \psi_r^{\bar d} \right] 
+ \mathcal L_{\rm top},
\eeq
where $\mathcal L_{\rm top}$ is the topological term
\beq
\mathcal L_{\rm top} = \frac{i\theta}{\pi} G_{a \bar b} \left( \p \varphi^a 
\bar \p \bar \varphi^{\bar b} - \bar \p \varphi^a 
\p \bar \varphi^{\bar b} \right),
\eeq
$G_{a \bar b}~(a, \bar b = 1, \cdots N-1)$ 
is the standard Fubini-Study metric
\beq
G_{a \bar b} = \frac{\p^2}{\p \varphi^a \p \bar \varphi^{\bar b}} 
\log \left( 1 + \sum_{c=1}^{N-1} |\varphi^c|^2 \right), 
\eeq
$\D$ and $\bar \D$ are pullbacks of the covariant derivative 
\beq
\D \psi_l^a = \p \psi_l^a + \Gamma_{bc}^a \p \varphi^b \psi_l^c, \hs{10}
\bar \D \psi_r^a = \bar \p \psi_r^a + \Gamma_{bc}^a 
\bar \p \varphi^b \psi_r^c, 
\eeq
and $\Gamma^a_{bc}$, $\bar \Gamma^{\bar a}_{\bar b \bar c}$ 
and $R_{a \bar b c \bar d}$ are the Christoffel symbol and 
curvature tensor 
\beq
\Gamma^a_{bc} = G^{\bar d a} \p_b G_{c \bar d}, \hs{5}
\bar \Gamma^{\bar a}_{\bar b \bar c} = G^{\bar a d} 
\p_{\bar b} G_{d \bar c}, \hs{5}
R^{\bar a}{}_{\bar b c \bar d} = \p_c \bar \Gamma^{\bar a}_{\bar b \bar d}.
\eeq
As in the case of the $\C P^1$ model, 
we impose the twisted boundary conditions
\beq
\varphi^a(y + 2 \pi R) = e^{2\pi i m_a R} \, \varphi^a(y), \hs{10} 
\psi_{l,r}^a(y+2\pi R) = e^{2\pi i m_a R} \, \psi_{l,r}^a(y), \label{eq:tbc-CPN}
\eeq
and introduce the following deformation potential 
which breaks the $\mathcal N = (2,2)$ supersymmetry 
\beq
\delta \mathcal L \ \equiv \ 
- \frac{\delta \epsilon}{N \pi R} \Delta \mu 
\ = \ \frac{\delta \epsilon}{\pi R} \left( \mu - \frac{1}{N} \sum_{a=1}^{N-1}m_a \right) , 
\hs{10}
\mu \equiv 
\sum_{a=1}^{N-1} \frac{m_a |\varphi^a|^2}{1+\sum_{c=1}^{N-1} |
\varphi^c|^2},
\eeq
where $\Delta$ is the Laplacian on the $\C P^{N-1}$
and $\mu$ is the moment map of the $U(1)$ symmetry 
used for the twisted boundary condition in Eq.~(\ref{eq:tbc-CPN}). 
Again, this deformation is inspired 
by the term induced by fermions in quantum mechanics
\cite{Fujimori:2017osz} . 
With this potential term, 
we can embed the single bion configuration $\varphi_B$ 
in Eq.~\eqref{eq:bion_ansatz} as a valley solution 
in the $\C P^{N-1}$ model. 
For example, $\varphi_B$ can be embedded 
into the $b$-th component as 
\beq
 \varphi^a = 0 ~~~(a \not = b), \hs{10}
\varphi^b = \varphi_B = e^{i m_b y} 
\left( e^{-\omega_b(x-x_-)-i\phi_-} 
+ e^{\omega_b(x-x_+) - i\phi_+} \right)^{-1},
\label{eq:bion_ansatz_cpn}
\eeq
where we have defined the parameter $\omega_b$ 
by replacing the parameter $\omega$ in \eqref{eq:omega_cp1} as 
\begin{equation}
\omega ~ \rightarrow ~ \omega_b \equiv m_b \sqrt{1 +\frac{g^2 \delta\epsilon}{\pi m_b R}}.
\label{eq:omega_cpn}
\end{equation}
As in the $\C P^1$ case, the valley solution satisfies 
the equation of motion of 
the deformed $\C P^{N-1}$ sigma model 
if the quasi-moduli parameters 
$x_\pm$ are adjusted to the values at the saddle points: 
for the real bion, for instance,  
\beq
x^{\rm rb}_\pm = x_0 \pm \frac{1}{2\omega_b} 
\log \frac{4\omega_b^2}
{\omega_b^{2}-m_b^{2}}, 
\hs{10}
e^{i \phi^{\rm rb}_\pm} = \pm e^{i\phi_0}.
\label{eq:bion_saddle_cpn}
\eeq 
The classical bion effective action $S_{\rm eff}(\eta)$ 
in the weak coupling limit \eqref{eq:deviation_from_saddle} 
now becomes 
\beq
S_{{\rm eff},b}(\eta) \ = \ \frac{4\pi m_b R}{g^2} 
\Big[ 1 - 2 \cos (\phi_+-\phi_-) e^{-m_b(x_+-x_-)} 
\Big] + 2 \delta\epsilon m_b |x_+-x_-| + {\mathcal O} \left( g^2 \right),
\label{eq:bare_eff_cpn}
\eeq
where we have assumed that 
$e^{-m_b(x_+-x_-)} \sim \mathcal O(g^2)$ 
as in the $\C P^1$ case.

Generalizing \eqref{eq:Z1Z2ratio} for the $\C P^1$ case, 
the single bion contribution of the $\C P^{N-1}$ model 
is given as a sum over the bion backgrounds as 
\beq
E_1 ~ \equiv \, - \lim_{\beta \rightarrow \infty} \frac{1}{\beta}  \frac{Z_1}{Z_0} ~ = \, - 2 \pi \sum_{b=1}^{N-1} 
\int_{\mathcal M_{r,b}} \! dx_r d \phi_r \, \mathcal J_b 
\, \exp \left( - S_{R,b} \right),
\label{eq:Z1Z2ratio_cpn}
\eeq
where $\mathcal J_b$ is 
the volume factor of the relative quasi-moduli space 
$\mathcal M_{r,b}$
and $S_{R,b}$ is the renormalized effective action 
including all the contributions from the fluctuations
\beq
S_{R,b} \ \equiv \ S_{{\rm eff}, b}
+\sum_{a}
\left(\left.\log\frac{\det' \! \Delta_B}{\det \Delta_{B}^0}\right|_a
- \left.\log \frac{\det \Delta_F}{\det \Delta_{F}^0}\right|_a
\right). 
\eeq
In the next subsections, 
we discuss the renormalization of the bion effective action 
due to the fluctuations around the bion background \eqref{eq:bion_ansatz_cpn}.

\subsection{One-loop determinants}
Let us consider the bosonic and fermionic fluctuations 
around the bion configuration in Eq.~\eqref{eq:bion_ansatz_cpn}. 
We can show that the fluctuations of the fields 
in the $b$-th component
$(\varphi^b,\psi_l^b,\psi_r^b)$ 
(the same component as the bion background)
 give the identical contribution as in the $\C P^1$ case. 
 
The one-loop determinant 
for the bosonic fluctuation $\delta \varphi^a~(a \not = b)$ 
can be calculated by the KK expansion. 
The contributions from the KK zero modes
have been calculated in Ref.\,\cite{Fujimori:2017osz}
\beq
\log \frac{\det' \! \Delta_B}{\det \Delta_B^0} \bigg|_{n=0,\,a} = ~ 
\log \left( \frac{m_a-m_b}{m_a} \right) -m_b x_r + \mathcal O(g^2). 
\label{eq:one-loop_zero}
\eeq
As shown in Appendix~\ref{appendix:cpn_oneloop}, 
the total contribution of the KK modes is given by
\beq
\sum_{n \not = 0} \log \frac{\det' \! \Delta_B}{\det \Delta_B^0} 
\bigg|_{n,\,a} 
= ~ \sum_{n=1}^{\infty} \Big[ X_{n,a} +  Y_{n,a} \cos \phi_r \, 
e^{-m_bx_r} + {\mathcal O}(g^2) \Big],
\label{eq:one-loop_nonzero}
\eeq
where $X_{n,a}$ and $Y_{n,a}$ are given by 
(see Appendices \ref{app:cpn_LO} and 
\ref{app:cpn_LKK}, respectively for details)
\beq
X_{n,a} \equiv  
\log \frac{\frac{n}{R}+(m_a-m_b)}{\frac{n}{R}-(m_a-m_b)} 
\frac{\frac{n}{R}-m_a}{\frac{n}{R}+m_a}, \hs{10}
Y_{n,a} \equiv \frac{4 m_b R}{n} + \mathcal O(n^{-2}),
\eeq
respectively.
The zeta function regularization gives 
\beq
\sum_{n=1}^{\infty} X_{n,a} &=& - 2 m_b R \log R \Lambda_0 - 
\log \frac{\Gamma\left( 1 + (m_a-m_b) R \right)}
{\Gamma\left(1-(m_a-m_b) R \right)} 
\frac{\Gamma\left( 1 - m_a R \right)}{\Gamma\left( 1 + m_aR \right)}, \\
\sum_{n=1}^{\infty} Y_{n,a} \hspace{2pt} &=& 4 m_b R \log R\Lambda_{0} + \cdots \,,
\eeq
where $\cdots$ denotes irrelevant terms 
in the weak-coupling limit.

The one-loop determinant of the fermionic fluctuations
can also be calculated by the KK decomposition. 
As in the $\C P^1$ case, 
the contributions from the non-zero modes cancel out 
and only the KK zero modes contribute to the determinant. 
In the weak coupling limit,  
the total contribution from the fermionic fluctuations 
is given by (see Appendix.\,\ref{app:cpn_F})
\beq
\sum_{a=1}^{N-1} \left.\log \frac{\det \Delta_F}{\det \Delta_{F}^0} \right|_a ~ = - N m_b x_r + {\mathcal O} (g^2). 
\label{eq:one-loop_fermion}
\eeq

\subsection{Contribution to partition function}
The one-loop determinants of the $b$-th component fields 
can be obtained from 
Eqs.\,(\ref{eq:logdKK})-(\ref{eq:logdKK2}) and
(\ref{eq:jacobian_cp1}) 
by replacing $m \to m_b$. 
Combining it with the classical bion effective action 
\eqref{eq:bare_eff_cpn} and the one-loop determinants of the 
$a$-th components ($a \not = b$) 
given in Eqs.\,\eqref{eq:one-loop_zero}, 
\eqref{eq:one-loop_nonzero} and \eqref{eq:one-loop_fermion},
we find that the integrand of the quasi-moduli integral 
\eqref{eq:Z1Z2ratio_cpn} is given by
 \beq
\!\!\!\!\!
\mathcal J_b \exp \left( - S_{R,b} \right) = \mathcal A_b \left( 
\frac{4 m_b^2 R}{g_R^2} \right)^2 \exp \! \left( 
- \frac{4 \pi m_b R}{g_R^2} \right) \exp \! \left( 
\frac{8 \pi m_b R}{g_R^2} \cos \phi_r \, e^{-m_b x_r} 
- 2 m_b \epsilon x_r \right) + \cdots, 
\label{eq:SRCPN}
\eeq
where we have defined the constant $\mathcal A_b$ as
\beq
\mathcal A_b ~ \equiv ~ \left[ \frac{\Gamma(1-m_b R)}
{\Gamma(1+m_b R)} \right]^2
\prod_{a \not = b} \frac{m_a}{m_a-m_b} 
\frac{\Gamma\left( 1 + (m_a-m_b)R \right)}
{\Gamma\left( 1 - (m_a-m_b)R \right)} 
\frac{\Gamma\left( 1 - m_a R \right)}{\Gamma\left( 1 + m_aR \right)}. 
\label{eq:coefficient}
\eeq
The parameter $\epsilon$ is 
the the effective deformation parameter
defined in the same way as in the $\C P^1$ model\footnote{
It would be helpful to comment that 
the linear term $-2m_b x_r$ in Eq.\,(\ref{eq:SRCPN}) emerges 
due to the bosonic KK zero modes 
in Eq.\,(\ref{eq:one-loop_zero}) and 
the fermionic KK zero modes in Eq.\,(\ref{eq:one-loop_fermion}) 
as $(N-2) m_b x_r - N m_b x_r = - 2m_b x_r$.}
\beq
\epsilon \, \equiv \, 1 + \delta\epsilon. 
\label{eq:epsilon_cpN}
\eeq
Note that $g_R^2$ in the renormalized effective action
Eq.\,(\ref{eq:SRCPN}) is the coupling constant
renormalized at $1/R$, 
which appeared as a result of the renormalization procedure. 
The renormalized effective action can also be written 
in terms of $\Lambda$ related to $g_R^2$ as
\beq
\frac{1}{g_R^2} ~= \, -\frac{N}{2\pi} \log |R \Lambda| \hs{10} \left( \ = ~ \frac{1}{g_0^2} -\frac{N}{2\pi} \log |R \Lambda_0| \right).
\eeq

Since the renormalized effective action in Eq.\,\eqref{eq:SRCPN} 
has the same form as that in the $\C P^1$ case, 
we can apply the Lefschetz thimble method to the 
quasi-moduli integral \eqref{eq:Z1Z2ratio_cpn} 
as in the previous case. 
Using the results of the thimble integral summarized in 
Appendix~\ref{app:LT}, 
we finally obtain the single bion contribution to 
the vacuum energy 
\beq
E_1 \ = \, -\frac{\Gamma(\epsilon)}{\Gamma(1-\epsilon)} 
e^{\mp \pi i \epsilon} \, 
\sum_{b=1}^{N-1} \, 2 m_b \mathcal A_b
\left( \frac{4\pi m_b R}{g_R^2} \right)^{2(1-\epsilon)} \!
|R \Lambda|^{2 N m_b R} + \cdots, 
\label{eq:CPN_singlebion}
\eeq
where the upper (lower sign) corresponds to 
the positive (negative) imaginary part given to $g_R^2$
in order to avoid the Stokes line in the parameter space. 
 Again, the single bion contribution vanishes 
when the deformation parameter $\delta \epsilon$ is 
turned off $(\epsilon = 1)$. 
This is consistent with the fact that 
the vacuum energy is exactly zero for a supersymmetric vacuum. 
A similar phenomenon can be seen 
in the model with $n_{\rm F}$ copies of 
fermionic degrees of freedom. 
For $n_{\rm F} = 1, 2, \cdots$, 
the one-loop determinant from the fermion fluctuations 
\eqref{eq:one-loop_fermion} 
is multiplied by $n_{\rm F}$
so that the result \eqref{eq:CPN_singlebion} 
can be easily generalized to the case of $n_{\rm F}$ copies of fermion by redefining $\epsilon$ as
\beq
\epsilon \, \equiv \, 1 + \delta\epsilon ~~ \rightarrow ~~ 
\epsilon \, \equiv \, 1 + \delta \epsilon + \frac{N}{2}( n_{\rm F} -1 ). 
\eeq
In the case of quantum mechanics, 
the general theorem states 
\cite{Kozcaz:2016wvy,Fujimori:2017osz} 
that the vacuum energy vanishes
when this $\epsilon$ becomes a positive integer. 
Also in the 2D case, Eq.~\eqref{eq:CPN_singlebion} 
implies that $E_1$ vanishes for $\epsilon \in \Z_{\geq 0}$.
Hence in the absence of the deformation $\delta \epsilon$, 
the single bion contribution vanishes 
for any $n_{\rm F}$ when $N$ is even.
On the other hand, when $N$ is odd, 
there is a non-trivial non-perturbative correction 
for even $n_{\rm F}$.
Such a case provides a simple example of
a non-trivial bion correction and resurgence structure
without the deformation parameter $\delta\epsilon$. 

When $E_1=0$ in the absence of the deformation, 
we need to consider expansion in powers of 
the deformation parameter $\delta \epsilon$ 
in order to see non-vanishing non-perturbative contributions
\beq
E_1 ~= \sum_{b=1}^{N-1} C_b |R \Lambda|^{2 N m_b R} \left[ \delta \epsilon + 2 \left(\psi(\epsilon_0)-\log\frac{4\pi m_b R}{g_R^2}\mp\frac{\pi i}{2}\right) \delta \epsilon^2 + \cdots \right], 
\label{eq:CPNE1}
\eeq
where $\psi(\epsilon_0)$ is the digamma function, 
$\epsilon_0 \equiv \epsilon \, |_{\delta \epsilon =0} \in \Z_{\geq 0}$ and 
\beq
C_b ~=\, -2 m_b \mathcal A_b
\left[  \, \Gamma(\epsilon_0) 
 \left(\frac{4\pi m_b R}{g_R^2}\right)^{1-\epsilon_0} \,
 \right]^2 .
\eeq
Although the imaginary ambiguity disappears 
at the leading order of the expansion, 
the higher order expansion coefficients have 
non-trivial imaginary parts with ambiguous signs. 
Since these ambiguities in the non-perturbative bion contribution
should be canceled by the Borel resummation ambiguity 
of the perturbation series, 
the large order behavior of the perturbation series can be 
determined from this single bion contribution 
as in the case of the $\C P^1$ model. 
 
For the $\Z_{N}$-symmetric boundary condition, 
which is realized by setting the parameters 
$m_a~(a=1,\cdots,N-1)$ as $m_a = \frac{a}{N} \frac{1}{R}$, 
the leading term of the result Eq.\,(\ref{eq:CPNE1}) 
is of order $|R\Lambda|^{2}$. 
This is consistent with the renormalon contribution 
$|\Lambda|^{2} \propto e^{-2\pi/(g_R^2 N)}$. 
This result again indicates that the renormalon ambiguity 
can be canceled by the bion contribution, 
and hence the bion could be identified as the infrared renormalon.


\section{Summary and discussion}
\label{sec:summary}
In this paper, we have calculated 
the semiclassical contributions 
from the bion saddle points 
in the ${\mathbb C}P^{N-1}$ models on 
${\mathbb R}\times S^{1}$ with twisted boundary conditions, 
with emphasis on its consistency with the infrared renormalon. 
We have discussed the bion contributions 
in the 2D ${\mathcal N}=(2,2)$ supersymmetric model 
and its non-supersymmetric generalization 
with $n_{\rm F}$ copies of fermions 
and the deformation parameter $\delta\epsilon$, 
including the cases corresponding to 
quasi-exactly-solvable model 
upon the dimensional reduction to quantum mechanics. 
We have derived the bion solutions
composed of a pair of fractional instanton and anti-instanton
by promoting that of the ${\mathbb C}P^{N-1}$ 
quantum mechanics to the 2D system. 
We discussed the renormalization of the effective action 
on the quasi-moduli space of the bion configurations, 
which is parametrized by the relative distance and phase 
between the fractional instanton and anti-instanton. 
The quantum fluctuation 
around the bion background 
renormalizes the coupling constant in the effective action, 
leading to the emergence of the dynamical scale. 
From the renormalized effective action, 
we obtained the bion contribution 
to the vacuum energy. 
Although the vacuum energy vanishes 
for the SUSY and quasi-exactly-solvable cases, 
we have shown that there are non-vanishing bion contributions 
exhibiting the structure expected from the resurgence theory 
by expanding the vacuum energy 
in powers of the deformation parameter $\delta \epsilon$.
We showed that the imaginary ambiguity 
in the bion contributions is consistent 
with the expected infrared renormalon ambiguity. 
This is the first result revealing 
the explicit relation between the bion contribution and 
the infrared renormalon ambiguity
in quantum field theories.

One of topics left for future works is the large-$N$ limit 
with the ${\mathbb Z}_{N}$ twisted boundary condition. 
Since the $\Z_{N}$ symmetric phase has been 
shown to be continuous \cite{Sulejmanpasic:2016llc} 
as the compactification radius is increased, 
it would be interesting to study whether 
the bion contribution survives in the large-$N$ limit. 
Studying the large radius limit will make 
the relation between the bion and the renormalon more direct. 

In this paper we have not considered 
the twisted masses for the chiral multiplets,
which can be introduced 
without breaking the $\mathcal N =(2,2)$ supersymmetry. 
The twisted masses  give a potential term proportional to 
a squared norm of a linear combination of 
the Killing vectors for the holomorphic isometries. 
Although there is no essential change in the single bion solutions,
such potential terms can modify the one-loop determinants.
It would be interesting to discuss 
how the bion contributions and imaginary ambiguities 
are modified in the presence of the twisted masses. 
 
We here comment on the full trans-series and 
complex multi-bion solutions. 
In the ${\mathbb C}P^{N-1}$ quantum mechanics, 
the multi-bion contributions are building blocks of 
the full trans-series of physical quantities. 
Such multi-bion solutions give 
non-perturbative contributions also 
in the present field theoretical case. 
However, we should remember that 
they are not enough 
in the 2D $\C P^{N-1}$ quantum field theory. 
In addition to them, 
there are bion configurations 
composed of instanton and anti-instanton
each of which has an integer topological charge. 
Such configurations also contribute to the full trans-series 
and it is quite possible that 
they may become more important 
as we increase the compactification radius $R$.  

Since the ${\mathbb C}P^1$ manifold can be embedded 
into any K\"ahler manifolds of the form of coset spaces $G/H$,
our work can be generalized to 
2D ${\cal N}=2$ SUSY nonlinear sigma models 
on K\"ahler $G/H$ manifolds
and their SUSY breaking deformations.
 
In 4D gauge theory 
such as Yang-Mills and QCD 
with an appropriate compactification, 
we may be able to take a similar procedure 
to derive contributions from bion configurations. 
One of the important questions in these theories is 
what are quasi-moduli and 
whether one can perform the quasi-moduli integral. 
Another question is whether bions are complex solutions 
of the complexified gauge theory.
Future works are devoted to the investigation on these questions.


\begin{acknowledgments}
The authors are grateful to the organizers and participants 
of ``RIMS-iTHEMS International Workshop on Resurgence Theory" 
at RIKEN, Kobe and ``Resurgent Asymptotics in Physics and Mathematics" 
at Kavli Institute for Theoretical Physics for giving them a 
chance to deepen their ideas. 
This work is supported by the Ministry of Education, Culture, 
Sports, Science, and Technology(MEXT)-Supported Program for the 
Strategic Research Foundation at Private Universities ``Topological 
Science" (Grant No. S1511006). This work is also supported in part 
by the Japan Society for the Promotion of Science (JSPS) 
Grant-in-Aid for Scientific Research (KAKENHI) Grant Numbers 
18K03627 (T.\ F.), 16K17677 (T.\ M.),  16H03984 (M.\ N.) and 
18H01217 (N.\ S., T.\ F., T.\ M. and M.\ N.). 
The work of M.N. is also supported in part 
by a Grant-in-Aid for Scientific Research on Innovative Areas 
``Topological Materials Science" (KAKENHI Grant No. 15H05855) 
from MEXT of Japan. 
The work of S.\ K. is supported in part by the US Department 
of Energy Grant No. DE-FG02-03ER41260.
\end{acknowledgments}

\appendix
\section{An example of quasi-moduli space}
\label{appendix:QMS}
In this appendix, we illustrate 
the concept of the quasi-moduli space 
by using an example of a simple zero dimensional model.
Let us consider the perturbation expansion 
of the following integral on $\R^2$
\beq
Z \equiv \int d^2 \phi \, \exp \left( -S/g^2 \right) ~~ \mbox{with} ~~
S \equiv S_0 + g^2 S_2, \hs{3} 
S_0 \equiv \Big[ (\phi^1)^2 + (\phi^2)^2 - 1 \Big]^2, \hs{3}
S_2 \equiv 2 \epsilon (\phi^1)^2. 
\eeq
The leading order part $S_0$ is invariant under the rotation 
and has degenerate minima 
corresponding to the spontaneously broken rotational symmetry
\beq
\frac{\p S_0}{\p \phi^i} \, \bigg|_{\phi = \varphi_0} = 0 
~~~ \Longrightarrow ~~~ 
\ba{c} \varphi_0^1 \\ \varphi_0^2 \ea \equiv \ba{c} \cos \eta \\ \sin \eta \ea.
\eeq
This flat direction is lifted by the symmetry breaking term $S_2$,
so that $\eta$ can be viewed as 
the pseudo-Nambu-Goldstone mode 
and only two points ($\eta = \pm \frac{\pi}{2}$) 
remain the discrete minima of $S$. 
Since the symmetry breaking term vanishes 
in the weak coupling limit $g \rightarrow 0$,
the parameter $\eta$ can also be viewed as the quasi-modulus, 
which becomes exact flat direction for $g = 0$. 
Let us determine the quasi-moduli space, 
which is described by the embedding 
$\eta \rightarrow \phi^i = \varphi^i(\eta)$
satisfying the valley equation\footnote{
This valley equation can be obtained 
from Eqs.\,(\ref{eq:valley}) and (\ref{eq:coef_valley}) 
for the ${\mathbb C}P^1$ model 
by replacing $G\to \delta_{ij}$ and 
$g^{\bar\beta\alpha} \to g^{\eta \eta} = 
1/(\partial_\eta \varphi^i)^2$. }
\beq
0 = P \ \ba{c} \displaystyle \p_{\phi^1} S \\ 
\displaystyle \p_{\phi^2} S \ea_{\phi=\varphi(\eta)} ,
\eeq
where $P$ is the projection operator onto 
the ``massive" direction
\beq
P \ = \ \mathbf 1_2 - \frac{1}{(\p_\eta \varphi^i)^2} 
\ba{c} \p_\eta \varphi^1 \\ \p_\eta \varphi^2 \ea \ba{cc} 
\p_\eta \varphi^1 & \p_\eta \varphi^2 \ea. 
\eeq
Starting from $\varphi_0^i(\eta)$, 
we can perturbatively solve the valley solution. 
Let $\xi^i(\eta)$ be the deviation from $\varphi_0^i(\eta)$
satisfying the orthogonality condition 
\beq
\ba{c} \varphi^1(\eta)\\ \varphi^2(\eta) \ea \ = \ 
\ba{c} \varphi_0^1(\eta) 
+ \xi^1(\eta) \\ \varphi_0^2(\eta) + \xi^2(\eta) \ea, \hs{10}
\xi_i \, \p_\eta \varphi_0^i = 0.
\eeq
Expanding $\xi^i$ as 
$\xi^i(\eta) = g^2 \xi^i_2(\eta) + g^4 \xi^i_4(\eta) + \cdots$, 
the valley equation can be solved order-by-order
\begin{align}
h \ba{c} \xi_2^1 \\ \xi_2^2 \ea  & 
= - 4 \cos^2 \eta \ba{c} \cos \eta \\ 
\sin \eta \ea \hs{-8} & 
\Longrightarrow ~ \ba{c} \xi_2^1 \\ \xi_2^2 \ea 
& = - \frac{1}{2} \epsilon \cos^2 \eta \ba{c} \cos \eta \\ 
\sin \eta \ea, \notag \\
h \ba{c} \xi_4^1 \\ \xi_4^2 \ea & = 
2 ( 3 \cos^2 \eta - 4 ) \cos^2 \eta \ba{c} \cos \eta \\ 
\sin \eta \ea \hs{-8} &
\Longrightarrow ~ \ba{c} \xi_2^1 \\ \xi_2^2 \ea 
& = \frac{1}{4} ( 3 \cos^2 \eta - 4 ) 
\cos^2 \eta \ba{c} \cos \eta \\ \sin \eta \ea, \notag \\
& & \vdots \hs{18} & \notag 
\end{align}
where $h$ is the Hessian of $S_0$
\beq
h = 8 \ba{cc} \cos^2 \eta & \sin \eta \cos \eta \\ \sin \eta \cos \eta 
& \sin^2 \eta \ea. 
\eeq
Actually, in the case of this simple example, 
we can exactly solve the valley equation
without expansion. 
By setting
\beq
\ba{c} \varphi^1(\eta)\\ \varphi^2(\eta) \ea \ 
= \ r(\eta) \ba{c} \cos \eta 
\\ \sin \eta \ea, 
\eeq
the valley equation can be rewritten as
\beq
0 \ = \ \frac{4 r X}{r^2 + (\p_\eta r)^2} 
\ba{c} ~ \p_\eta ( r \sin \eta ) 
\\ - \p_\eta (r \cos \eta ) \ea, ~~ ~\mbox{with} ~~~
X \equiv g^2 \epsilon \cos \eta \, \p_\eta( r \sin \eta ) + r(r^2-1).  
\eeq
The solution of $X=0$ satisfying 
$\lim_{g \rightarrow 0} r = 1$ is an ellipse 
\beq
r(\eta) = \sqrt{\frac{1 - g^2 \epsilon}{1-g^2 \epsilon \sin^2 \eta}}, 
\hs{10} 
\left( \frac{1}{1-g^2 \epsilon} (\varphi^1)^2 + (\varphi^2)^2 
= 1 \right).
\eeq
On this quasi-moduli space, the effective action is given by
\beq
S_{\rm eff}(\eta) \ = \ S \, |_{\phi = \varphi(\eta)} \ 
= \ \big[ r(\eta)^2 
- 1 \big] ^2 - 2(1-g^2 \epsilon) \big[ r(\eta)^2 - 1 \big]. 
\eeq
By changing the valuables from 
$(\phi^1, \phi^2)$ to $(\eta,\delta \varphi)$
\beq
\ba{c} \phi^1 \\ \phi^2 \ea = 
\ba{c} \varphi^1(\eta) \\ \varphi^2(\eta) \ea 
+ \frac{g \, \delta \varphi}{\sqrt{(\p_\eta \varphi^1)^2
+(\p_\eta \varphi^2)^2}} 
\ba{c} \phantom{-} \p_\eta \varphi^2 \ \\ 
- \p_\eta \varphi^1 \ \ea, 
\eeq
the original integral can be rewritten as
\beq
Z =  \int_0^{2\pi} d \eta \, \sqrt{g_{\eta \eta}} \, 
\exp \left( - \frac{S_{\rm eff}}{g^2} - S_q \right),
\eeq
where $S_q$ is given by
\beq
\exp \left( - S_q \right) = g \int  d\delta \varphi \, 
\frac{\sqrt{\tilde g_{\eta \eta}}}{\sqrt{g_{\eta \eta}}} 
\exp \left( - \frac{S_{\rm fluc}}{g^2} \right), \hs{10} 
S_{\rm fluc} \equiv S - S_{\rm eff}, 
\eeq
with
\beq
\frac{\sqrt{\tilde g_{\eta \eta}}}{\sqrt{g_{\eta \eta}}} \ 
= \ \frac{\sqrt{(\p_\eta \phi^1)^2+(\p_\eta \phi^2)^2}}
{\sqrt{(\p_\eta \varphi^1)^2+(\p_\eta \varphi^2)^2}} \ 
= \ 1 + \frac{g \, \delta \varphi}{1-g^2 \epsilon} 
\left( \frac{r^2}{\sqrt{r^2+(\p_\eta r)^2}} \right)^3. 
\eeq
Thanks to the definition of the valley solution, 
there is no linear term of $\delta \varphi$ in $S_{\rm fluc}$ 
\beq
\frac{S_{\rm fluc}}{g^2} \ = \ \delta \varphi^2 \left[  
\left( g \, \delta \varphi + 
\frac{2r^2}{\sqrt{r^2 + (\p_\eta r)^2}} \right)^2 
+ \frac{2 r^2(\p_\eta r)^2}{r^2+(\p_\eta r)^2} \right] \ 
= \ 4 \delta \varphi^2 + \mathcal O(g),
\eeq
so that we can integrate out $\delta \varphi$ by expanding 
the integrand in powers of $g^2$ as in the standard perturbation 
theory 
\beq
\exp \left( - S_q \right) \ = \ \frac{\sqrt{\pi} g}{2} 
\left[ 1 + \frac{g^2 \epsilon}{2} \cos^2 \eta + \mathcal O(g^4) \right].
\eeq
Then the integrand for $Z$ can be expanded as
\beq
Z \ = \ \frac{\sqrt{\pi} g}{2} \int_0^{2\pi} d\eta \, \left[ 1 
+ g^2 \epsilon^2 \cos^4 \eta + \mathcal O(g^4) \right] 
\exp \left( - 2 \epsilon \cos^2 \eta \right). 
\eeq
Evaluating the quasi-moduli integral, 
we obtain the perturbation series of $Z$
\beq
Z \ = \ \pi^{\frac{3}{2}} g \, e^{-\epsilon} I_0(\epsilon) 
\bigg[ 1 + \frac{g^2 \epsilon}{4} \left( 2 \epsilon - ( 1 + 2 \epsilon) 
\frac{I_1(\epsilon)}{I_0(\epsilon)} \right) + \mathcal O(g^4) \bigg], 
\eeq
where $I_0(\epsilon)$ denotes 
the modified Bessel function of the first kind.
We can check that the perturbation series derived in this way
is consistent with that which can be obtained from 
the Borel resummed form of $Z$
\beq
Z = \frac{\pi}{2} \int_0^\infty dt \, e^{-\frac{t}{g^2}} \, \frac{
f \left( 1+\sqrt{t} \right) + f \left( 1-\sqrt{t} \right)}{\sqrt{t}}, 
\hs{10}
f(z) = e^{-\epsilon z} I_0(\epsilon z),
\eeq
which can be obtained 
by the change of variables $(\phi^1,\phi^2) = r(t) \, 
(\cos \eta, \sin \eta)$ with
\beq
r(t) = \left\{ 
\begin{array}{cc} 
\sqrt{1 + \sqrt{t}} & \mbox{for $|\phi^i|^2 > 1$} \\ 
\sqrt{1 - \sqrt{t}} & \mbox{for $|\phi^i|^2 < 1$} 
\end{array} 
\right.,
\eeq
and the $\eta$ integration. 

\section{One-loop determinants 
around single bion background in $\C P^1$ model}
\label{appendix:cp1_det}
In this appendix, we calculate the one-loop determinant 
around the single bion ansatz \eqref{eq:bion_ansatz} 
in the $\C P^1$ model. 
For simplicity, 
we fix the center of mass position and overall phase  
and set $x_{\pm} = \pm x_r/2$ and $\phi_{\pm} = \pm \phi_r/2$. 
We will use the following theorem to calculate functional determinants
(see, e.g., Appendix B of Ref.\,\cite{Fujimori:2016ljw}). 
Let $\xi^\pm$ be functions such that
\beq
(- \p_x^2 + V ) \, \xi^\pm = 0, \hs{10} 
\lim_{x \rightarrow \pm \infty} \xi^\pm = e^{\mp M x}, \hs{10}
(M^2 \equiv  V( x \rightarrow \pm \infty)).
\eeq
Then the functional determinant of $- \p_x^2 + V$ 
is given by
\beq
\frac{\det (- \p_x^2 + \hs{1} V \hspace{5pt} )}{\det ( - \p_x^2 + M^2 )}
~= \, \lim_{x \rightarrow \mp \infty} e^{\pm M x} \xi^\pm. 
\label{eq:det_theorem}
\eeq

\subsection{Bosonic one-loop determinant in the KK decomposition}
\label{app:cp1_LO}
Let us first consider the functional determinant of 
$\Delta_B$ defined in Eq.\,\eqref{eq:DeltaB}. 
Since the background is independent of $y$ 
except for the twist factor, 
it is convenient to use the KK expansion 
for the bosonic fluctuation
\beq
\xi = \sum_{n=-\infty}^\infty \xi_n(x) \, 
e^{i \left( \frac{n}{R} + m \right) y}. 
\eeq
Then the functional determinant $\det \Delta_B$ decomposes 
into an infinite product of the contributions 
from the KK modes $\det \Delta_B |_n$.
Since the contribution from the KK zero mode 
is essentially the same as the 1D case \cite{Fujimori:2016ljw},
we focus on the non-zero mode $(n \not = 0)$ in the following.

As explained in Sec.\,\ref{subsec:eff_action}, 
we consider the weak coupling limit 
keeping the background bion ansatz 
in the vicinity of the saddle points.
In other words, we fix the deviation from the saddle point 
\beq
\delta x_r \, \equiv \, x_r - \frac{1}{\omega} \log \frac{4 \omega^2}{\omega^2 - m^2},
\eeq
and take the weak coupling limit $g \rightarrow 0$. 
In this limit, the operator $\Delta_B$ in Eq.~(\ref{eq:DeltaB}) 
becomes a diagonal matrix, 
i.e. the equations for $\xi$ and $\bar \xi$ 
become independent.  
For the $n$-th KK mode of $\xi$, 
the leading order part of $\Delta_B$
is given by
\beq
\Delta_B|_n = \left\{
\begin{array}{lll} 
\Big[ \p_x + \frac{n}{R} - m \tanh (m d_-) \Big] 
\Big[ \p_x - \frac{n}{R} + m \tanh (m d_-) \Big] & + ~ \mathcal O(g^2) \phantom{\bigg]} & \mbox{for $x \approx - x_r/2$} \\
\Big[ \p_x + \frac{n}{R} - m \Big] 
\Big[ \p_x - \frac{n}{R} + m \Big] & + ~ \mathcal O(g^2) ~~ & \mbox{for $x \, \approx \, 0$} \\
\Big[ \p_x - \frac{n}{R} - m \tanh (m d_+) \Big] 
\Big[ \p_x + \frac{n}{R} + m \tanh (m d_+) \Big] & + ~ \mathcal O(g^2) \phantom{\bigg]} & \mbox{for $x \approx  x_r/2$}
\end{array} 
\right.,
\eeq
where $d_{\pm}$ are the distances 
from the constituent fractional instantons
\beq
d_{\pm} \equiv x \mp \frac{x_r}{2}. 
\eeq
By applying the theorem \eqref{eq:det_theorem}, 
we can determine $\det \Delta_B |_n$,  
from the solution of 
the differential equation $\Delta_B|_n \, \xi = 0$. 
Solving the equation in each region, 
we obtain the solution as 
\beq
\xi =\left\{ 
\begin{array}{lll} 
\frac{1}{2} \, e^{\frac{n}{R} x - \frac{m x_r}{2}} \, {\rm sech}(m d_-) \bigg[ a_1 - a_2 F_-(x) \bigg] & + ~ \mathcal O(g^2) \phantom{\bigg]} & \mbox{for $x \approx - x_r/2$} \\
b_1 \, e^{\left( \frac{n}{R}-m \right)x} + b_2 \, e^{-\left( \frac{n}{R}-m \right) x} \phantom{\bigg[} & + ~ \mathcal O(g^2) ~~ & \mbox{for $x \, \approx \, 0$} \\
\frac{1}{2} \, e^{-\frac{n}{R} x - \frac{m x_r}{2}} \, {\rm sech} (md_+) \bigg[ c_1  + c_2  F_+(x) \bigg] & + ~ \mathcal O(g^2) \phantom{\bigg]} & \mbox{for $x \approx  x_r/2$}
\label{eq:linear_sol}
\end{array} 
\right.,
\eeq
where $F_\pm(x)$  are given by 
\beq
F_\pm(x) &=& 8 \left( \frac{n}{R} + m \right) e^{m x_r} \int dx \, e^{\pm \frac{2n}{R}x} \cosh^2 ( m d_\pm).  
\eeq
Let us consider the KK modes with $n > 0$.  
The solution $\xi^-$ which degreases exponentially 
for $x \rightarrow -\infty$
can be obtained by setting $a_1=1, ~a_2=0$ 
in the general solution \eqref{eq:linear_sol}. 
Then, by connecting the solutions in the neighboring regions, 
we can determine the coefficients $b_1$ and $c_2$ as
\beq
b_1 = e^{-mx_r}, \hs{10} 
c_2 = \frac{\frac{n}{R}-m}{\frac{n}{R}+m} e^{-2m x_{r}} .
\eeq
From the asymptotic form of $\xi_-$ for large $x$, 
we find that $\det \Delta_B |_n$ for $n > 0$ is given by
\beq
\frac{\det \Delta_B}{\det \Delta_{B}^0} \bigg|_{n} = \, 
\lim_{x \rightarrow \infty} e^{- \left( \frac{n}{R} + m \right) x} \, \xi^- 
\ = \ c_2 
\ = \ \frac{\frac{n}{R}-m}{\frac{n}{R}+m} e^{-2m x_{r}}.
\label{eq:det_positive_n}
\eeq

For $n<0$, the solution $\xi^-$ 
which degreases exponentially 
for $x \rightarrow -\infty$
can be obtained by setting $a_1=0,~ a_2=1$.
Connecting the solution, we can determine $b_2$ and $c_1$ as
\beq
b_2 = \frac{\frac{n}{R} + m}{\frac{n}{R}-m} e^{mx_r}, \hs{10}
c_1 = \frac{\frac{n}{R} + m}{\frac{n}{R}-m} e^{2m x_{r}}. 
\eeq
Therefore, $\det \Delta_B |_n$ for $n < 0$ is given by
\beq
\frac{\det \Delta_B}{\det \Delta_{B}^0} \bigg|_{n} = \, 
\lim_{x \rightarrow \infty} e^{\left( \frac{n}{R} + m \right) x} \xi^- 
\ = \ c_1 
\ = \ \frac{\frac{n}{R}+m}{\frac{n}{R}-m} e^{2m x_{r}}.
\label{eq:det_negative_n}
\eeq
Combining the contributions 
from the positive and negative KK modes in 
Eqs.~\eqref{eq:det_positive_n} and \eqref{eq:det_negative_n}, 
we find that the total contribution of bosonic KK modes is given by
\beq
\sum_{n=1}^\infty \left[ 
\log \frac{\det \Delta_B}{\det \Delta_{B}^0} \bigg|_{n} 
+ \log \frac{\det \Delta_B}{\det \Delta_{B}^0} \bigg|_{-n} \right] \ = \ 
2 \sum_{n=1}^\infty \log \frac{\frac{n}{R}-m}{\frac{n}{R}+m}. 
\eeq
Since this infinite sum is divergent, 
let us consider the zeta function regularization
\beq
\sum_{n=1}^\infty \log \frac{\frac{n}{R}-m}{\frac{n}{R}+m} &=& 
\lim_{s \rightarrow 0} \frac{\p}{\p s} 
\left[ 
\sum_{n=1}^\infty 
\left( \frac{\Lambda_0}{\frac{n}{R} + m} \right)^{s} 
- \sum_{n=1}^\infty \left( \frac{\Lambda_0}{\frac{n}{R} - m} \right)^{s}
\right],
\label{eq:zeta_regularization}
\eeq
where $\Lambda_0$ is an arbitrary parameter 
which can be identified with a UV cutoff scale.
Using the Hurwitz zeta function $\zeta (s,z)$, which satisfies
\beq
\zeta(s,z) = \sum_{n=0}^\infty \frac{1}{(z+l)^s}, \hs{10}
\zeta(0,z) = -z + \frac{1}{2}, \hs{10} 
\lim_{s \rightarrow 0} \frac{\p}{\p s} \zeta (s,z) = \log \frac{\Gamma(z)}{\sqrt{2\pi}}, 
\eeq
we obtain the following regularized KK mode contribution 
with the cutoff dependence 
\beq
\sum_{n=1}^\infty \log \frac{\frac{n}{R}-m}{\frac{n}{R}+m} &=& \lim_{s \rightarrow 0} \p_s \Big[ (R \Lambda_0)^s \Big\{ \zeta(s,1+mR) - \zeta(s, 1-mR) \Big\} \Big] \notag \\
&=& - 2 m R \, \log R \Lambda_0 + \log \frac{\Gamma (1+ m R)}{\Gamma(1-mR)},
\label{eq:logdKKapp}
\eeq
This result gives the first equality in Eq.~(\ref{eq:logdKK2}).

\subsection{Large KK momentum expansion}
\label{app:cp1_LKK}
Here, we discuss the UV divergence of 
the bosonic one-loop determinant in more detail 
by using the large KK momentum expansion. 
For large KK momentum (large $n$), 
the bosonic one-loop determinant can be expanded as
\beq
\frac{\det \Delta_B}{\det \Delta_B^0} \bigg|_n 
= 1 + \frac{1}{n} A + \mathcal O(n^{-2}).
\eeq
The expansion coefficient $A$ 
determines the UV divergent part as
\beq
\sum_{n=1}^\infty \log \frac{\det \Delta_B}{\det \Delta_B^0} \bigg|_n  = \, A \hspace{1pt} \log R \Lambda_0 + \{\mbox{UV finite terms}\},
\label{eq:divergent_part}
\eeq
where we have used the relation
\beq
\sum_{n=1}^\infty \frac{1}{n} = \log R \Lambda_0 + \gamma ,
\label{eq:zeta}
\eeq
which can be obtained by differentiating Eq.\,\eqref{eq:logdKKapp}
with respect to $m$ and setting $m=0$. 
Eq.\,\eqref{eq:divergent_part} implies that 
if we are interested only in the UV divergent part, 
it is sufficient to calculate the constant $A$ 
by using the large KK momentum expansion. 

By expanding the fluctuation as
\beq
\xi(x,y) = \sum_{n=-\infty}^\infty \xi_n(x) \, e^{i \left(\frac{n}{R} 
+ m \right) y}, \hs{10}
\bar \xi(x,y) = \sum_{n=-\infty}^\infty \tilde \xi_n(x) \, 
e^{i \left(\frac{n}{R} - m \right) y},
\eeq
the equation for the bosonic fluctuations become
\beq
\Delta_B \ba{c} \xi \\ \bar \xi \ea ~ \rightarrow ~ 
\Delta_{B} |_n 
\ba{c} \xi_n \\ \tilde \xi_{n} \ea,
\eeq
where $\Delta_{B} |_n$ is the operator which can be obtained by
replacing $\nabla$, $\bar \nabla$, $\nabla^\ast$ and 
$\bar \nabla^\ast$ in Eq.\,\eqref{eq:DeltaB} as
\beq
\nabla ~ \rightarrow \frac{1}{2} \left( \p_1 + \frac{n}{R} + m \right) + 2 i A_z , \hs{10}
\bar \nabla \,\, \rightarrow \frac{1}{2} \left( \p_1 - \frac{n}{R} - m \right)+ 2 i A_{\bar z}, \\
\nabla^\ast \rightarrow \frac{1}{2} \left( \p_1 + \frac{n}{R} - m \right) - 2 i A_z , \hs{10}
\bar \nabla^\ast \rightarrow  \frac{1}{2} \left( \p_1 - \frac{n}{R} + m \right) - 2 i A_{\bar z}.
\eeq
By generalizing the theorem \eqref{eq:det_theorem}, 
the functional determinant of $\Delta_B |_n$ 
can be calculated as follows 
(see Appendix B of Ref.\,\cite{Fujimori:2016ljw}). 
For $n>0$, let $\Xi_n$ be a 2-by-2 matrix 
(a linearly independent pair of 
solutions of the linearized equation) such that
\beq
\Delta_{B} |_n \, \Xi_n = 0, \hs{10} 
\Xi_n =\ba{cc} e^{\left( \frac{n}{R} + m \right) x} & 0 \\ 0 
& e^{\left( \frac{n}{R} - m \right) x} \ea \exp W_n,
\eeq 
where $W$ is a 2-by-2matrix which converges in the limit 
$x \rightarrow -\infty$
\beq
\lim_{x \rightarrow - \infty} W_n = {\rm const}.
\eeq
Then the one-loop determinant can be written as
\beq
\log \frac{\det \Delta_B}{\det \Delta_{B}^0} \bigg|_n ~=~ 
\int_{-\infty}^{\infty} dx \, \p_x \log \det \exp W_n 
~=~ \int_{-\infty}^{\infty} dx \, \tr \, \p_x W_n. 
\eeq
By expandong $W_n$ in powers of $1/n$ as
\beq
W_n = W_{n,0} + \frac{1}{n} W_{n,1} + \frac{1}{n^2} W_{n,2} + \cdots,  
\eeq
we can recursively determine $\p_x W_{n,k}$ 
by solving $\Delta_B |_n \, \Xi_n = 0$ order-by-order. 
Then we can show that
\beq
\int_{-\infty}^{\infty} dx \, \tr \, \p_x W_n ~= \, - \frac{2R}{n} \int_{-\infty}^{\infty} dx \, \frac{|\p_x \varphi_B|^2 + m^2 |\varphi_B|^2}{(1+|\varphi_B|^2)^2} + \mathcal O(n^{-2}). 
\label{eq:determinat_integral}
\eeq
The KK modes with $n<0$ gives the same contribution 
to the divergent part. 
As in the case of the classical bion effective action
(see footnote \ref{footnote:seff}), 
we can evaluate the integral in Eq.\,\eqref{eq:determinat_integral} 
up to the subleading order in the weak coupling limit. 
By using the zeta function regularization in Eq.~\eqref{eq:zeta}, 
we find that 
\beq
\sum_{n=1}^\infty \left[ \log \frac{\det \Delta_B}{\det \Delta_{B}^0} \bigg|_n + \log \frac{\det \Delta_B}{\det \Delta_{B}^0} \bigg|_{-n} \right] \ = \, 
- 4 m R \log R \Lambda_0 \, \Big[ 1 - 2 \cos \phi_r \, e^{-m x_r} + \mathcal O(g^4) \Big] + \cdots. 
\eeq
This correctly renormalizes the coupling constant $g$ 
in the effective action. 
From this expression, 
we can read the second equality in Eq.~(\ref{eq:logdKK2}).

\subsection{Fermionic one-loop determinant}
\label{app:cp1_F}
Next, let us consider the fermionic one-loop determinant
in the bion background.
For a single pair of $(\chi_l, \chi_r)$ in Eq.\,\eqref{eq:S0}, 
the one-loop determinant is given by
\beq
\det \Delta_F  = \det (\nabla \bar \nabla) = \det (\bar \nabla \nabla), 
\hs{10}
\det \Delta_F^0 = \det (\p \bar \p). 
\eeq
Note that both $\nabla$ and $\bar \nabla$ defined in Eq.\,\eqref{eq:covariant_derivative1} 
have no zero mode in the bion background. 
It is convenient to expand the fluctuations into the KK modes as
\beq
\chi_{l}(x,y) \, = \, e^{-2i \int^x dx A_x } \sum_{n=-\infty}^{\infty} 
e^{i M_n y} \chi_{l,n}(x), \hs{5}
\chi_{r}(x,y) \, = \, e^{-2i \int^x dx A_x } \sum_{n=-\infty}^{\infty} 
e^{i M_n y} \chi_{r,n}(x),
\eeq
where $M_n$ is the KK mass 
\beq
M_n = \frac{n}{R} + m. 
\eeq
Since each KK sector is an eigen subspace of the operators 
$\Delta_F$ and $\Delta_F^0$, 
we can decompose the determinants as 
\beq
\log \frac{\det \Delta_F}{\det \Delta_{F}^0} = 
\sum_{n=-\infty}^\infty \log \frac{\det \Delta_F}{\det \Delta_{F}^0} 
\bigg|_{n}. 
\eeq
In the $n$-th KK sector, 
the explicit form of the operators are given by
\beq
\nabla \bar \nabla \big|_n = \frac{1}{4} \left(\p_x + 2 A_y 
+ M_n \right) \left(\p_x - 2 A_y - M_n \right), \hs{10}
\p \bar \p = \frac{1}{4} \left( \p_x^2 - M_n^2 \right). 
\eeq
Let us calculate the determinant 
by using the theorem \eqref{eq:det_theorem}. 
Let $\chi_{\pm}$ be the solutions of 
$\nabla \bar \nabla \big|_n \, \chi_{\pm} = 0$
with the following asymptotic behaviors
\beq
\chi_{\pm} \rightarrow \left\{ 
\begin{array}{cc} e^{ \mp \left| M_n \right| x} & \mbox{for 
$x \rightarrow \pm \infty$} \\ 
\mathcal C^{\pm} e^{ \mp \left| M_n \right| x} + \mathcal D^{\pm} 
e^{ \pm \left| M_n \right| x} & \mbox{for $x \rightarrow \pm \infty$} 
\end{array} \right..
\eeq
Then the ratio of the determinants is given by
\beq
\frac{\det (\nabla \bar \nabla) }{\det \, (\p \bar \p) \, } \bigg|_n 
= \, \mathcal C^+ = \, \mathcal C^-.
\eeq
The solution of $\bar \nabla \chi = 0$ 
gives $\chi_-$ for $n \geq 0$ and $\chi_+$ for $n < 0$  
\beq
\chi_- &=& e^{M_n x} \exp \left[ 2 \int_{-\infty}^x dx' \, 
A_y(x') \right] ~~~(\mbox{for $n \geq 0$}), \\
\chi_+ &=& e^{M_n x} \exp \left[ 2 \int_{~ \infty}^x \, dx' \, 
A_y(x') \right] \hspace{1pt} ~~~(\mbox{for $n < 0$}). 
\eeq
Therefore the determinant is given by
\beq
\frac{\det (\nabla \bar \nabla) }{\det \, (\p \bar \p) \, } \bigg|_n 
= \exp \left[ \pm 2 \int_{-\infty}^\infty dx \, A_y(x) \right] 
\hs{5} \mbox{(for $n \geq 0$ and $n < 0$ respectively)} .
\eeq
It follows that all the fermionic contributions cancel out except for 
the KK zero mode
\beq
\log \frac{\det \Delta_F}{\det \Delta_{F}^0} ~=~ 
\sum_{n=-\infty}^\infty \log \frac{\det (\nabla \bar \nabla) }{\det \, (\p \bar \p) \, }
\bigg|_{n} 
~=~ 2 \int_{-\infty}^\infty dx \, A_y(x) 
~= - 2 m x_{r} + \mathcal O(g^2). 
\label{eq:fermion_determinant}
\eeq
This result gives Eqs.~(\ref{eq:DeltaFKK}) and (\ref{eq:jacobian_cp1}).

\section{One-loop determinants around single bion background in 
$\C P^{N-1}$ model}
\label{appendix:cpn_oneloop}

In this appendix, we discuss the one-loop determinants 
around the single bion backgrounds in the $\C P^{N-1}$ model 
in Eq.~\eqref{eq:bion_ansatz_cpn}. 
As in the $\C P^1$ case, 
we fix the center of mass position and overall phase  
and set $x_{\pm} = \pm x_r/2$ and $\phi_{\pm} = \pm \phi_r/2$.

\subsection{Bosonic one-loop determinant in the KK decomposition}
\label{app:cpn_LO}
When the single bion ansatz of the $\C P^1$ model 
is embedded in the $b$-th component field $\varphi^b$, 
the fluctuation of $\varphi^b$
gives the same determinant as in the $\C P^1$ case in 
Eqs.\,(\ref{eq:logdKK})-(\ref{eq:logdKK2}) and (\ref{eq:jacobian_cp1}) 
with $m$ replaced by $m_b$. 
For $\delta \varphi^a~(a \not = b)$, 
it is convenient to redefine the fields as 
\beq
\delta \varphi^a = g \sqrt{1+|\varphi_B|^2} \, 
\exp \left( - i \int^x dx' A_x(x') \right) \, \xi^a, 
\eeq
and expand the normalized fluctuation $\xi^a$ into the KK modes
\beq
\xi^a = \sum_{n=-\infty}^{\infty} e^{i \left( \frac{n}{R} 
+ m_a \right) y} \, \xi_n^a. 
\eeq
Then the linearized equation 
for the $n$-th KK mode of $\xi^a$ becomes
\beq
0 ~=~ \Delta_B |_{n,a} \, \xi^a_n ~=~ 
\Bigg[ \p_x^2 - \left( \frac{n}{R} + m_a + A_y \right)^2 
+ \frac{|\p_x \varphi_B|^2 
+ m_b^2 |\varphi_B|^2}{(1+|\varphi_B|^2)^2} \Bigg] \, \xi_n^a 
+ \mathcal O(g^2). 
\label{eq:linearized_cpn}
\eeq
The leading order part of the operator $\Delta_B |_{n,a}$ 
is given by 
\beq
\Delta_B|_{n,a} = \left\{
\begin{array}{lll} 
\Big[ \p_x + M_{n,a} - \frac{m_b}{1+e^{-2m_b d_-}} \Big] 
\Big[ \p_x - M_{n,a}+ \frac{m_b}{1+e^{-2m_b d_-}} \Big] & + ~ \mathcal O(g^2) \phantom{\bigg]} & \mbox{for $x \approx - x_r/2$} \\
\Big[ \p_x + M_{n,a}- m_b \Big] 
\Big[ \p_x - M_{n,a} + m_b \Big] & + ~ \mathcal O(g^2) ~~ & \mbox{for $x \, \approx \, 0$} \\
\Big[ \p_x - M_{n,a} + \frac{m_b}{1+e^{2m_b d_+}} \Big] 
\Big[ \p_x + M_{n,a} - \frac{m_b}{1+e^{2m_b d_+}} \Big] & + ~ \mathcal O(g^2) \phantom{\bigg]} & \mbox{for $x \approx  x_r/2$}
\end{array} 
\right.,
\eeq
where $M_{n,a}$ is the KK mass and 
$d_{\pm}$ are the distances from the fractional instantons
\beq
M_{n,a} = \frac{n}{R} + m_a, \hs{10} 
d_{\pm} = x \mp \frac{x_r}{2}.
\eeq
The leading order solution of $\Delta_B |_{n,a} \, \xi_n^a$ 
can be obtained by connecting the solutions in the neighboring regions. 
For $n \geq 0$, the decreasing solution 
as $x \rightarrow - \infty$ is given by
\beq
\xi_n^a ~=~ e^{M_{n,a} \hspace{1pt} x} \times \left\{ 
\begin{array}{ll} 
(1+e^{2 m_b d_-})^{-\frac{1}{2}} \phantom{\bigg[} & \mbox{for $x \approx -x_r/2$} \\
\exp \left[  - m_b d_- \right] + \cdots \phantom{\bigg[} & \mbox{for $x \approx 0$} \\
(1+e^{-2 m_b d_+})^{-\frac{1}{2}}  \left( \frac{M_{n,a}-m_b}{M_{n,a}} e^{-m_b x_r} + e^{-2m_b x} \right) + \cdots  \phantom{\bigg[} & \mbox{for $x \approx x_r/2$}
\end{array} \right..
\eeq
Similarly, for $n < 0$ 
the decreasing solution 
as $x \rightarrow \infty$ is given by
\beq
\xi_n^a ~=~ e^{M_{n,a} \hspace{1pt} x} 
\left\{ 
\begin{array}{ll} 
(1+e^{2 m_b d_-})^{-\frac{1}{2}} \frac{M_{n,a}}{M_{n,a}-m_b} e^{m_b x_r} + \cdots \phantom{\bigg[} & \mbox{for $x \approx - x_r/2$} \\
\frac{M_{n,a}}{M_{n,a}-m_b} e^{-m_b d_+} + \cdots \phantom{\bigg[} & \mbox{for $x \approx 0$} \\
(1+e^{-2 m_b d_+})^{-\frac{1}{2}} \left( 1 + \frac{M_{n,a}}{M_{n,a}-m_b} e^{-2m_b d_+} \right) \phantom{\bigg[} & \mbox{for $x \approx x_r/2$}
\end{array} \right..
\eeq
From the asymptotic behavior of these solutions, 
the one-loop determinant can be read off as
\beq
\log \frac{\det \Delta_B}{\det \Delta_B^0} \bigg|_{n,a} \ \approx \ \mp \left[  \log \frac{\frac{n}{R} + m_a}{\frac{n}{R} + m_a-m_b} + m_b x_r \right],
\eeq
where $-$ is for $n \geq 0$ and $+$ is for $n < 0$. 
Therefore, the contribution of the $a$-th component of the bosonic fluctuation to the one-loop determinant is given by
\beq
\log \frac{\det \Delta_B}{\det \Delta_B^0}\bigg|_{a} \ 
= \ \sum_{n=1}^\infty \left[ \log \frac{\frac{n}{R} + m_a-m_b}
{\frac{n}{R} - m_a+m_b} - \log \frac{\frac{n}{R} + m_a}
{\frac{n}{R} - m_a} \right] - \log \frac{m_a}{m_a-m_b} - m x_r.
\eeq
By using the zeta function regularization in Eq.~\eqref{eq:zeta_regularization},
we find 
\beq
\log \frac{\det \Delta_B}{\det \Delta_B^0}\bigg|_{a} \, = \,
 - 2 m_b R \, \log R \Lambda_0 - \log \frac{\Gamma (1+ (m_a-m_b) R)}
{\Gamma(1- (m_a-m_b) R)}\frac{\Gamma(1-m_a R)}{\Gamma (1+ m_a R)} 
\frac{m_a}{m_a-m_b} - m x_r . 
\eeq
This gives Eq.\,\eqref{eq:one-loop_zero} and the first term of Eq.\,(\ref{eq:one-loop_nonzero}).

\subsection{Large KK momentum expansion}
\label{app:cpn_LKK}
To see the UV divergence in more detail, 
let us consider the large KK momentum expansion. 
The solution of the linearized equation \eqref{eq:linearized_cpn} 
which decreases as $x \rightarrow -\infty$ 
can be obtained by setting
\beq
\xi^a_n = \exp \left[ \pm \left( \frac{n}{R} + m_a \right) x 
+ \sum_{k=0}^\infty \frac{1}{n^k} W_{n,a,k} \right],
\eeq
and expanding the equation in powers of $n$. 
Here $+$ is for $n \geq 0$ and $-$ is for $n<0$. 
We can show that $\p_x W_{n,a,k}$ satisfy
\beq
\p_x W_{n,a,0} = \pm A_y , \hs{5}
\p_x W_{n,a,1} = \mp \frac{R}{2} \left[ \frac{|\p_x \varphi_B|^2 
+ m_b^2 |\varphi_B|^2}{(1+|\varphi_B|^2)^2} 
+ m_b \p_x \left( \frac{1}{1+|\varphi_B|^2} \right) \right] , ~~~ \cdots.
\eeq
By using the theorem \eqref{eq:det_theorem}, 
we find that 
the total contribution of the KK modes is given by
\beq
\log \frac{\det \Delta_B}{\det \Delta_B^0} \bigg|_{a,{\rm KK}}
= ~ \sum_{k=0}^\infty \sum_{n=1}^{\infty} \int_{-\infty}^{\infty} dx \, 
\left[ \frac{1}{n^k} \p_x W_{n,a,k} + \frac{1}{(-n)^k} \p_x W_{-n,a,k} \right].
\eeq
For $k=1$, the summation with respect to $n$ is divergent. 
By applying the zeta function regularization \eqref{eq:zeta},
we find that the divergent part is given by 
\beq
\log \frac{\det \Delta_B}{\det \Delta_B^0} \bigg|_{a,{\rm KK}}
&=& - R \log R \Lambda_0 \int_{-\infty}^{\infty} dx \frac{|\p_x \varphi_B|^2 + m_b^2 |\varphi_B|^2}{(1+|\varphi_B|^2)^2} + \cdots \notag \\
&=& - 2 m_b R \log R \Lambda_0 \Big[ 1 - 2 \cos \phi_r e^{-m_b x_r} + \mathcal O(g^4) \Big] + \cdots. \phantom{\bigg]}
\eeq
This divergence consistently renormalizes 
by the coupling constant in the classical bion effective action. 
From this result, we can read off 
the second term of Eq.~(\ref{eq:one-loop_nonzero}).

\subsection{Fermionic one-loop determinant} 
\label{app:cpn_F}
Next, let us calculate the fermionic one-loop determinant 
for a single pair of fermions $(\psi_l^a, \psi_r^a)$. 
It is convenient to redefine the fermionic fields as
\beq
\psi_{l,r}^a \ = \ g (1+|\varphi_B|^2)^{\frac{q_a}{2}} \chi^a_{l,r}, \hs{5}
\mbox{with} ~~
q_a = \left\{ 
\begin{array}{ll} 
2 & \mbox{for $a=b$} \\
1 & \mbox{for $a \not =b$} 
\end{array} \right.. 
\eeq
Then the linearized equations for the fermionic fluctuations 
become
\beq
\left( \p + i q_a A_z \right) \chi_l^a = 0, \hs{10} 
\left( \bar \p + i q_a A_{\bar z} \right) \chi_r^a = 0,
\eeq
The fermionic one-loop determinant 
can be calculated in an analogous way 
as in the case of the $\C P^1$ model. 
Generalizing the formula \eqref{eq:fermion_determinant} 
to each component, 
we find that the fermionic contributions 
to the determinant is given by
\beq
\sum_{a=1}^{N-1} \left.\log \frac{\det \Delta_F}{\det \Delta_{F}^0} \right|_a 
~ = \ \sum_{a=1}^{N-1} q_a \int_{-\infty}^\infty dx \, A_y(x) 
~ = \, - N m_b x_{r} + \mathcal O(g^2).
\eeq
This gives the fermionic one-loop determinant 
in Eq.~(\ref{eq:one-loop_fermion}).

\section{Lefschetz thimble integral}
\label{app:LT}

In this appendix, we summarize the procedure 
to evaluate the quasi-moduli integral 
by means of the Lefschetz thimble method 
following the argument of Ref.\,\cite{Fujimori:2016ljw}. 
The quasi-moduli integrals discussed in this paper take the form
\beq
[{\mathcal I}\bar{\mathcal I}] \ \equiv \ \int_{-\infty}^{\infty} d x_r 
\int_{-\pi}^{\pi} d \phi_r \, e^{-S}, \hs{5}
S(x_r,\phi_r) \ \equiv \, - \frac{8\pi m R}{g_R^2} \cos \phi_r \, e^{-m x_r} + 2 m \epsilon x_r + \mathcal O (g^2) \,. 
\label{eq:QMI}
\eeq
We first complexify $g_R^2$ as $g_R^2 \to g_R^2 e^{i\theta}$
to avoid the Stokes line.
The variables $x_r$ and $\phi_r$ 
are also complexified as 
$x_r = x_{R}+i x_{I} \in {\mathbb C}$, 
$\phi_r =\phi_{R}+i \phi_{I} \in {\mathbb C}$.
By solving the equations 
$\p_{x_r} S = 0$ and $\p_{\phi_r} S = 0$, 
we find that the saddle points are labeled 
by an integer $\sigma \in \Z$ 
\begin{align}
x_{\sigma} \ = \ \frac{1}{m} \log\left( \frac{4 \pi mR}{\epsilon g_R^2} \right) + \frac{i}{m} (\sigma \pi - \theta), \hs{10} 
\phi_{\sigma} \ = \, - (\sigma - 1)\pi ~~~ ({\rm mod} \, 2\pi) \,,
\label{eq:critical}
\end{align}
where $\sigma = 0$ and $\sigma = \pm 1$ 
corresponds to the real and complex bions, respectively. 
The thimbles $\mathcal J_\sigma$ and 
the dual thimbles $\mathcal K_\sigma$ associated 
with these saddle points are obtained 
by solving the flow equations
\beq
\frac{d x_{r}}{dt} = \frac{1}{2m} \overline{\frac{\partial S \,}
{\partial x_{r}}}, \hs{10}
\frac{d \phi_{r}}{dt} = \frac{m}{2} \overline{\frac{\partial S \,}
{\partial \phi_{r}}},
\eeq
where the coefficients in the right hand sides of these equations 
are determined by the metric of the quasi-moduli space. 
By solving these equations, 
we find that the thimbles $\mathcal J_\sigma$ are 
the planes specified by
\beq
m x_I = \sigma \pi - \theta, \hs{10} \phi_R = - (\sigma-1) \pi\,,
\eeq
while the dual thimbles $\mathcal K_\sigma$ are specified by the equations 
\begin{align}
m x_R - \phi_I  =
\log \left[ \frac{4 \pi mR}{\epsilon g_R^2}
\frac{\sin \mathcal Y}{\mathcal Y} \right], \hs{10}
m x_R + \phi_I =
\log \left[ \frac{4 \pi mR}{\epsilon g_R^2}
\frac{\sin \tilde{\mathcal Y}}{\tilde{\mathcal Y}} \right], 
\end{align}
where $\mathcal Y$ and $\tilde{\mathcal Y}$ are given by
\beq
\mathcal Y \equiv m x_I + \phi_R -\pi+\theta, \hs{5}
\tilde{\mathcal Y} \equiv m x_I - \phi_R -(2\sigma -1)\pi +\theta, \hs{5} - \pi \leq \mathcal Y \leq \pi, \hs{3} - \pi \leq \tilde{\mathcal Y} \leq \pi .
\eeq
By looking into these thimbles and dual thimbles, 
we find that the intersection numbers $(n_{-1},n_{0},n_{1})$
of the original integration contour 
and the dual thimbles $\mathcal K_{-1}$, $\mathcal K_1$ and $\mathcal K_0$ are given by
\beq
(n_{-1} \, , \, n_{0} \, , \, n_{1}) ~=~
\left\{ 
\begin{array}{cc}
(\, 0 \, , \, -1\, , \, 1 \,) ~&~ \mbox{for $\theta > 0$} \\
(\, -1 \, , \, 1 \, , \, 0 \,) ~&~ \mbox{for $\theta < 0$} 
\end{array} 
\right..
\eeq
Therefore, in the limit $\theta \rightarrow \pm 0$, 
the integral \eqref{eq:QMI} has the ambiguity depending 
on the sign of $\theta$
\beq
[{\mathcal I}\bar{\mathcal I}] ~=~ 
\left\{ 
\begin{array}{cc}
Z_{\sigma=1} - Z_{\sigma=0} \, ~~&~ \mbox{for $\theta \rightarrow +0$} \\
Z_{\sigma=0} - Z_{\sigma=-1} ~&~ \mbox{for $\theta \rightarrow -0$} 
\end{array} 
\right.,
\eeq
where $Z_\sigma$ denotes the integral 
along the thimble $\mathcal J_\sigma$ 
\beq
Z_\sigma \ = \ \int_{\mathcal J_\sigma} dx_r d \phi_r \, \exp \left[ - S(x_r, \phi_r) \, \right] 
\ = \ \frac{i}{2m} \left(  {4\pi m R\over{g_R^{2}} } 
\right)^{-2\epsilon} e^{-2 \pi i \epsilon \sigma} \, 
\Gamma \left( \epsilon \right)^2.
\eeq
Therefore, the integral \eqref{eq:QMI} is evaluated as
\beq
[{\mathcal I}\bar{\mathcal I}] ~=~ 
\frac{1}{m}\left(  {4\pi m R\over{g_R^{2}} } 
\right)^{-2\epsilon} 
\sin \epsilon \pi \, \Gamma \left( \epsilon \right)^2 \times
\left\{ 
\begin{array}{ll}
e^{-\pi i \epsilon} ~&~ \mbox{for $\theta=+0$} \\
e^{+\pi i \epsilon} ~&~ \mbox{for $\theta=-0$}
\end{array} 
\right..
\eeq


\end{document}